# Water storage capacity of the Martian mantle through time

Junjie Dong[12]*, Rebecca A. Fischer[1], Lars P. Stixrude[3], Carolina R. Lithgow-Bertelloni[3], Zachary T. Eriksen[1], Matthew C. Brennan[1]

[1] Department of Earth and Planetary Sciences, Harvard University, Cambridge, Massachusetts 02138, USA.
[2] Department of the History of Science, Harvard University, Cambridge, Massachusetts 02138, USA.
[3] Department of Earth, Planetary, and Space Sciences, University of California, Los Angeles, California 90095, USA.

*Corresponding author: Junjie Dong (junjiedong@g.harvard.edu)

**Keywords:** Mars, interior; water storage capacity; nominally anhydrous minerals

**Highlights**
- We present depth-dependent water storage capacity profiles for the solid Martian mantle.
- The bulk $H_2O$ storage capacity of the present-day Martian mantle is ~9 km GEL.
- The bulk $H_2O$ storage capacity of the initial Martian mantle was ~5 km GEL.

**Abstract**
Water has been stored in the Martian mantle since its formation, primarily in nominally anhydrous minerals. The short-lived early hydrosphere and intermittently flowing water on the Martian surface may have been supplied and replenished by magmatic degassing of water from the mantle. Estimating the water storage capacity of the solid Martian mantle places important constraints on its water inventory and helps elucidate the sources, sinks, and temporal variations of water on Mars. In this study, we applied a bootstrap aggregation method to investigate the effects of iron on water storage capacities in olivine, wadsleyite, and ringwoodite, based on high-pressure experimental data compiled from the literature, and we provide a quantitative estimate of the upper bound of the bulk water storage capacity in the FeO-rich solid Martian mantle. Along a series of areotherms at different mantle potential temperatures ($T_p$), we estimated a water storage capacity equal to $9.0^{+2.8}_{-2.2}$ km Global Equivalent Layer (GEL) for the present-day Martian mantle at $T_p = 1600$ K and $4.9^{+1.7}_{-1.5}$ km GEL for the initial Martian mantle at $T_p = 1900$ K. The water storage capacity of the Martian mantle increases with secular cooling through time, but due to the lack of an efficient water recycling mechanism on Mars, its actual mantle water content may be significantly lower than its water storage capacity today.

## 1 Introduction

Unlike Earth, which has vast oceans on its surface, Mars is primarily cold and dry today. The majority of its surface water at present likely lies in the polar-layered deposits (PLD) and in shallow ground ice. Melting the PLD and ground ice reservoirs would produce an $H_2O$ layer with an estimated thickness of 20–40 m if it were spread evenly over the Martian surface (Carr and Head, 2015), referred to as a Global Equivalent Layer (GEL). Early Mars may have had a much larger surface water reservoir, as suggested by its geomorphology (e.g., Carr, 1996) and abundance of ancient clay minerals (e.g., Ehlmann and Edwards, 2014). The high D/H ratio of the present Martian atmosphere (at least 5–7 times higher than the terrestrial value, e.g., Villanueva et al., 2015) can also be interpreted as the result of loss of early surface water reservoirs (Chassefière et al., 2013; Kurokawa et al., 2014; Scheller et al., 2021). The current size of the Martian surface water reservoir is the result of a complex history of crustal hydration, atmospheric escape, and mantle outgassing since its formation (Jakosky, 2020).

Water, which has been stored in the Martian mantle rocks since its formation, can lower their viscosity and affect the thermal evolution of the Martian interior, which further alters melting in the mantle and hence controls crustal formation as well as water outgassing onto the Martian surface (e.g., Hauck and Phillips, 2002; Fraeman and Korenaga, 2010; Grott et al., 2011; Morschhauser et al., 2011; Sandu and Kiefer, 2012; Ruedas et al., 2013a, 2013b; Breuer et al., 2016; Ogawa, 2016). For a better understanding of how the water budget of the Martian surface and interior reservoirs have evolved through geologic time, quantitative constraints on water in the Martian mantle are indispensable.

Current geochemical estimates of the Martian mantle water content are often based on our existing Martian meteorite archives, and their estimated water contents fall in the range 14–250 ppm wt $H_2O$ (summarized in Filiberto et al., 2019, and references therein). However, these meteorite-based estimates have two major drawbacks (Filiberto et al., 2019): (1) the water content derived from one or a few meteorite samples may only represent their source regions instead of the average mantle; and (2) several secondary processes, including shock, magmatic degassing upon cooling, and terrestrial contamination may have altered the original water contents of these meteorites. Alternatively, as observational constraints on Martian mantle structure continue to improve (e.g., Mocquet and Menvielle, 2000; Civet and Tarits, 2014; Johnson et al., 2020; Ruedas and Breuer, 2021, Stähler et al., 2021), the mantle water content may be constrained by geophysical observables such as seismic velocities and attenuation and electrical conductivity in the future (Verhoeven and Vacher, 2016; Ruedas and Breuer, 2021).

In this study, we present a mineral physics approach to put constraints on the amount of water in the Martian mantle. On a microscopic level, water is stored in the form of structurally-bound hydroxyl groups (OH) in the nominally anhydrous minerals (NAMs), which comprise the bulk of the Martian mantle. Instead of being infinite water reservoirs, individual mantle NAMs have finite OH solubilities, also known as their water storage capacities. Numerous mineral physics studies (cf. Dong et al., 2021, and references therein) have investigated the water storage capacities in major NAMs common to the mantles of terrestrial planets. The experimental data on the major NAMs' water storage capacities available in the literature were used in this study to model the bulk water storage capacity of the Martian mantle based on its stable mineral assemblages (Tables S11–S13). An analogous calculation for the Earth indicates that the terrestrial mantle today has an estimated bulk water storage capacity of ~2.3 times the modern surface ocean mass (Dong et al., 2021). The Martian mantle is expected to be primarily composed of olivine (*ol*) and its high-pressure polymorphs, wadsleyite (*wd*) and ringwoodite (*rw*), as well as pyroxene (*px*) and garnet (*gt*), similar to Earth's mantle at the same pressures (Bertka and Fei, 1997). However, the bulk water storage capacity of the Martian mantle could be quite different from that of the Earth, due to the smaller planetary size of Mars, the different Martian mantle potential temperature, and the potentially more iron-rich composition of the Martian mantle. Here we first quantified the effects of iron on the water storage capacities of major mantle NAMs, and then built a bulk water storage capacity model for the solid Martian mantle and tracked its evolution along mantle adiabats as a function of $T_p$.

## 2 Methods

### 2.1 Equilibrium phase assemblages of the Martian mantle

We computed a series of adiabatic temperature gradients (areotherms) of a simplified Martian mantle composition as a function of pressure for $T_p$ from 1500 to 1900 K, using the thermodynamic code HeFESTo and the self-consistent parameter set contained therein (Stixrude and Lithgow-Bertelloni, 2005, 2011) (Fig. S1). The equilibrium mineral assemblages along these adiabats were calculated to constrain the mineralogical evolution of the Martian mantle with secular cooling over geologic time (Fig. S2–S6). Conversion from pressure to depth was performed using a thin shell calculation (Supplementary Text S1). The simplified bulk silicate Mars (BSM) composition in our model was a six-component system of $Na_2O$–$CaO$–$FeO$–$MgO$–$Al_2O_3$–$SiO_2$, with abundances primarily based on Taylor (2013) ("T13", Table S1). In addition, a pseudosection for the T13 composition was calculated between 3 and 22 GPa and between 1500 K and 2500 K to explore how Martian mantle equilibrium mineral assemblages vary with pressure ($P$) and temperature ($T$). We also included another four BSM compositions, from Liebske and Khan (2019) ("LK19", Table S2–S3), Yoshizaki and McDonough (2010) ("YM20", Table S4) and Khan et al. (2022) ("K22", Table S5) to test the sensitivity of our calculations to mantle composition (Fig. S2–S6, Section 3.2). At least 25 mantle phases (e.g., $ol$, $wd$, $rw$) with 47 endmember species (e.g., forsterite–fayalite) are likely to be stable over the $P$–$T$ conditions of terrestrial planetary mantles and have been considered here (cf. Table A1 in Stixrude and Lithgow-Bertelloni, 2011). The phase stability predictions of HeFESTo are generally consistent with the experimental results based on a BSM composition similar to T13 (Bertka and Fei, 1997, Khan et al., 2021). For the full description of the thermodynamic method of HeFESTo, see Stixrude and Lithgow-Bertelloni (2005, 2011).

### 2.2 Bulk $H_2O$ storage capacity of Martian mantle rocks

In calculating the bulk water storage capacity of the solid Martian mantle, the water storage capacities of individual NAMs at each pressure and temperature were estimated and summed. The water storage capacities of the major NAMs in the Martian mantle ($ol$, $wd$, and $rw$) can be parameterized as:

$$\ln\left(c_{H_2O}\right) = a + \frac{n}{2} \cdot \ln f_{H_2O}(P, T) + \frac{b + c \cdot P + d \cdot X_{Fe}}{T} \quad (1)$$

with $T$ in kelvin (K), $P$ in gigapascals (GPa), $f_{H_2O}$ (water fugacity) in GPa, and $X_{Fe}$ (iron content) in mole fraction (see more details in Dong et al., 2021). The constants $a$, $b$, and $c$ are related to the changes in entropy, enthalpy, and volume of the hydration reaction, respectively; $d$ expresses the effect of iron (as FeO); and $n$ is the fugacity exponent. We did not include the effects of oxygen fugacity, $f_{O_2}$, because its influence on the water storage capacities of $ol$, $wd$, and $rw$ is likely negligible (*ol*: Withers and Hirschmann, 2008; Gaetani et al., 2014; *wd*: Druzhbin et al., 2021; *rw*: Fei et al., 2020a). The presence of trace elements may enhance water incorporation in NAMs such as olivine at low pressures (<3 GPa) and low temperatures (<900°C) (Padrón-Navarta and Hermann, 2017). However, the effects of trace elements diminish at the adiabatic Martian mantle conditions we are interested in here, and hence were not included in the calculation. For simplification, we merged two constant terms, $\ln\frac{1}{2n}$ and *a*, in Equation 1 of Dong et al. (2021) into the constant *a* (Equation 1) of this study.

To fit Equation 1 for each NAM, we used a compilation of high-pressure mineral physics data on NAM water storage capacities (in *ol*, *wd*, and *rw*) from Dong et al. (2021), in which all water measurements based on Fourier transform infrared spectroscopy (FTIR) were corrected to the same calibration (*ol*: Withers et al., 2012; *wd/rw*: Bolfan-Casanova et al., 2018). For this study, we made several improvements to the Dong et al. (2021) data compilation: (1) a number of iron-rich samples ($X_{Fe}$ = 0.19–0.44) have been added to investigate how water storage capacities in NAMs may be different from terrestrial values in the FeO-rich mantle of Mars. The newly-added data can be found in Tables S11–13 of this study. (2) We excluded experimental data obtained from complex rock compositions (e.g., a peridotitic or pyrolitic composition) because the water storage capacity model is thermodynamically-derived for individual NAM phases. (3) We excluded experimental data with <1 wt% initial water content for olivine and those with <5 wt% for wadsleyite and ringwoodite to avoid water-undersaturated experiments. (4) We excluded experimental data for olivine that equilibrated with excess ferropericlase, (Mg,Fe)O, because a) the water storage capacities in these excess (Mg,Fe)O olivine samples are likely different from those equilibrated with excess $SiO_2$ at low pressures (Fei et al., 2020b), and hence they contribute additional uncertainty to the analysis; and b) the run products of these excess (Mg,Fe)O experiments do not resemble the mineral assemblages of the upper mantles of either Earth or Mars.

The bulk water storage capacity of Martian mantle rocks, $c_{H_2O}^{mantle}$, can be approximated by a weighted average (by their proportions *X*) of the water storage capacities of individual stable mantle phases at each *P–T*, including olivine or its polymorphs (*i*) and the coexisting mantle NAMs (*j*):

$$c_{H_2O}^{mantle}=\left(c_{H_2O}^{NAM}\right)_i \cdot \left(X_i+\sum_j\left(X_j \cdot D_{H_2O}^{i/j}\right)\right) \quad (2)$$

where $D_{H_2O}^{i/j}$ is the water partition coefficient between *i* and *j*, $\left(c_{H_2O}^{NAM}\right)_i$ is the water storage capacity of olivine or its polymorphs, and $\left(c_{H_2O}^{NAM}\right)_j=\left(c_{H_2O}^{NAM}\right)_i \cdot D_{H_2O}^{i/j}(T,P,X_{Al})$ is the water storage capacity of coexisting mantle NAMs. We used several literature compilations of uncorrected water storage capacity data (Hauri et al., 2006; Keppler & Bolfan-Casanova, 2006) to calculate $D_{H_2O}^{i/j}$ from the quotient of their uncorrected water storage capacities, $\frac{\left(c_{H_2O}^{NAM}\right)_i}{\left(c_{H_2O}^{NAM}\right)_j}$. Because the water storage capacity data for olivine (*i*, numerator) and its coexisting mantle NAMs, orthopyroxene, clinopyroxene, and garnet (*j*, denominator), require similar correction factors, the quotient $D_{H_2O}^{i/j}$ is virtually unaffected regardless of whether the numerator and denominator are corrected or not, as long as the treatment is consistent. The effects of iron on the water partition coefficients (but not the water storage capacities) are negligible compared with the effects of aluminum (Withers et al., 2011) and hence are not included in the parameterizations. Details of the parameterizations for these partition coefficients as a function of temperature, pressure, and aluminum content, $X_{Al}$, can be found in Dong et al. (2021). In addition, individual NAM water storage capacities are not strictly additive for complex rocks. Solid mantle rocks may have lower storage capacities than the sums of their individual minerals (e.g., Ardia et al., 2012; Tenner et al., 2012). Thus, the calculated water storage capacities for the solid Martian mantle mineral assemblages, $c_{H_2O}^{mantle}$, represent upper bounds.

## 2.3 The effect of iron on solid mantle $H_2O$ storage capacity

In our previous study that focused on Earth's mantle (Dong et al., 2021), we explored the effects of iron on the water storage capacities of *ol*, *wd*, and *rw*, but with an experimental data compilation that only included NAMs with $X_{Fe} < 0.19$. We found that the compositional dependence $d$ in Equation 1 was statistically insignificant (p > 0.01) for all experimental data with $X_{Fe} < 0.19$. However, the significance threshold for p-values (p = 0.01) is established by convention (Wasserstein and Lazar, 2016), and this finding for $X_{Fe} < 0.19$ (Dong et al., 2021) does not necessarily mean an absence of iron effects in these minerals at higher $X_{Fe}$. A $\frac{d \cdot X_{Fe}}{T}$ term with a p-value slightly greater than 0.01 only indicates that the iron effect cannot be readily resolved due to the resolution of the data for $X_{Fe} < 0.19$.

The Martian mantle is typically estimated to be rich in iron, with a bulk silicate Mg# ($=100 \times X_{Mg}/[X_{Mg}+X_{Fe}]$, molar) of ~75 (e.g., Taylor, 2013) compared to the bulk silicate Earth's Mg# of ~89 (e.g., Workman and Hart, 2005). The range of $X_{Fe}$ (in mole fraction) calculated for major NAMs throughout the Martian mantle (*ol*, *wd*, and *rw*) span approximately 0.2–0.4 for the T13 bulk composition (depending on specific $P$–$T$ conditions as well as the mineral assemblages along adiabats with $T_p$ = 1600–1900 K; Fig. S2–S6).

From a thermodynamic point of view, the Gibbs free energy of NAM hydration may be approximated as a linear function of $X_{Fe}$ (Schmalzried, 1995):

$$\Delta G(P, T, X_{Fe}) \sim \Delta G^*(P, T) + X_{Fe}\, \Delta G^*(P,T)'\ (3)$$

where $X_{Fe}\, \Delta G^*(P, T)'$ represents the deviation of the Gibbs free energy of hydration from that of the Mg endmember due to the presence of iron. The water concentration $c_{H_2O}$ is expected to change exponentially with $X_{Fe}$ (Zhao et al., 2004):

$$c_{H_2O} \propto \exp\left(-\Delta G^*(P, T)' \cdot \frac{X_{Fe}}{RT}\right)\ (4)$$

Therefore, the effects of iron on the water storage capacities of *ol*, *wd*, and *rw* may become significant at $X_{Fe}$ = 0.2–0.4, even though they are currently not resolvable for $X_{Fe} < 0.19$ using a robust nonlinear regression (Dong et al., 2021). In the context of an iron-rich Martian mantle for this study, we included additional experimental data with $X_{Fe} > 0.19$ from the literature (Table

S11–S13) to fit the $\frac{d \cdot X_{Fe}}{T}$ term (Equation 1) and to extrapolate the water storage capacities of NAMs to $X_{Fe} = 0.2$–0.4. Previous experiments demonstrate that the water storage capacity of *ol* increases by a factor of ~2–3 as its $X_{Fe}$ is increased from 0.2 to 0.4 at 3–6 GPa (Withers et al., 2011), so at least for olivine, iron has noticeable effects on water storage capacity at $X_{Fe} > 0.2$. However, no experimental study has systematically investigated the effects of iron on the water storage capacities of *wd* and *rw* with $X_{Fe} > 0.2$. If we directly fit Equation 1 to the compiled experimental dataset with the $\frac{d \cdot X_{Fe}}{T}$ term, the scarcity and imbalance of iron-rich samples, in particular for *wd* and *rw*, may cause an overfitting of *d*.

To avoid overfitting *d*, while also incorporating all experimental measurements into our estimate of the Martian bulk mantle water storage capacity and its uncertainty, we applied a statistical method called "bootstrap aggregation" or "bagging" (Fig. 1). The "bagging" method averages a large number (*N*) of the bulk mantle water storage capacity models, $M_1$, $M_2$, ..., $M_n$. These models are predicted based on different bootstrapped/resampled datasets and their regressions of Equation 1, with each model having a variance of $\sigma^2$. This method reduces variance embedded in the fitted $X_{Fe}$ coefficients by generating a model-averaged prediction, $\bar{M}$, with a variance of $\sigma^2/N$ (James et al., 2013). In other words, we quantified the uncertainty in the $\frac{d \cdot X_{Fe}}{T}$ terms based on experimental data including only limited data at high iron contents ($X_{Fe} > 0.19$), and then we incorporated the uncertainties associated with the extrapolations in $X_{Fe}$ directly as part of the uncertainty in the bulk mantle water storage capacity.

For the "bagging" method, we first resampled the literature data with replacement and created $10^4$ copies of the bootstrapped dataset (Step 1 in Fig. 1). We then fit a separate set of water storage capacity models (for each of *ol*, *wd*, and *rw*; Equation 1) to each resampled copy of the dataset (Step 2 in Fig. 1), resulting in $10^4$ mantle water storage capacity profiles along each adiabat for $T_p = 1500$–1900 K (Step 3 in Fig. 1). All regressions were unweighted due to the lack of inter-laboratory standard practices of error analyses. Finally, we aggregated the $10^4$ water storage capacity profiles as a function of depth for each $T_p$ and obtained the best-fit profiles by model averaging (mean) (Step 4 in Fig. 1). We integrated each water storage capacity profile as a function of depth to obtain the bulk mantle water storage capacity for each $T_p$. To quantify the uncertainty associated with this model, we report the mean bulk mantle water storage capacities as the best-fit values, along with medians as well as 90/10, 95/5 and 99/1 confidence intervals in Table S6. We use the 95/5 confidence intervals as the error bars throughout the text for consistency. This "bagging" method, unlike the Monte Carlo method used in Dong et al. (2021), considers all of the existing iron-rich NAM samples without presupposing the effects of iron. The trade-off is that the "bagging" method does not allow us to report a single set of regression parameters for each NAM explicitly, because the bulk mantle water storage capacity is based on the average of $10^4$ regressions for each NAM ($3 \times 10^4$ regressions in total). Despite this different approach, the average effects of pressure and temperature found by these regressions are consistent with those in Dong et al. (2021) (Fig. S7).

In addition, the uncertainties in the models of the Martian internal structure (e.g., crust thickness, core radius) and mantle chemistry (e.g., bulk silicate Mars composition) are not included in the error analysis. The bulk silicate Mars composition from Taylor (2013) (based on a volatile-depleted CI chondritic composition and element correlations in Martian meteorites) was used as the primary Martian mantle composition in this study. The effects of using other bulk silicate Mars compositions are discussed in Section 3.2. The crustal thickness and the core–mantle boundary (CMB) depth were assumed to be 50 km (e.g., Knapmeyer-Endrun et al., 2021) and 1600 km (e.g., Stähler et al., 2021), respectively, based primarily on InSight data as well as recent Martian internal structure models (e.g., Brennan et al., 2020).

## 3 Results and discussion

### 3.1 Depth-dependent profiles of the solid Martian mantle $H_2O$ storage capacity and the effects of Fe

Using this bootstrapping approach, we found that iron generally increases the water storage capacity of *ol* (Fig. 2a). The uncertainty in the regression increases at high pressures and temperatures due to a scarcity of experimental data at higher $P$–$T$ conditions. For *wd*, we found that the resampling of the experimental data typically produces a positive effect of iron on the *wd* water storage capacity; however, a significant number of regressions produce either negative or no effect of iron on the *wd* water storage capacity (Fig. 2b), which is due to inconsistencies in water measurements in the available literature data on *wd*. Fei and Katsura (2021) show that at similar temperatures, the water storage capacity in Fe-bearing *wd* is indeed higher than that in Fe-free *wd*, but they also noted that there is no experimentally-resolvable effect of iron within the Fe-bearing samples with Mg# > 88, consistent with the regression results in Dong et al. (2021). This relatively high variation in the extrapolation of the iron effect for *wd* is well characterized in the $10^4$ regressions of bootstrapped datasets (Fig. 2b) and is taken into account in the estimates of the bulk solid mantle water storage capacity in Section 3.2. For *rw*, we found that the regressions of bootstrapped datasets suggest a slightly negative effect of iron on water storage capacity, which is in good agreement with experimental data at 1900–2000 K (Fei and Katsura, 2020a; Fig. 2c).

We computed $10^4$ $P$–$T$ water storage capacity diagrams based on a representative set of bootstrapped regression parameters. Aggregating these results by model averaging, we constructed a best-fit water storage capacity $P$–$T$ diagram for the Martian mantle (Fig. 3a) as well as water storage capacity profiles along its adiabats (Fig. 4a–c). We found that the maximum water storage capacity along Martian adiabats is often reached in *wd* within the *ol*–*wd* transition region (0.5–1.3 wt%, Fig. 4a–c), where the lower temperatures and higher $X_{Fe}$ in *wd* give rise to a higher water storage capacity (iron preferentially partitions into *wd* within the *ol*–*wd* transition region, Fig. S2). Other abrupt changes in the profiles are consequences of phase transformations in the Martian mantle. The relatively large uncertainties in the extrapolation of the iron effect on the *wd* water storage capacity (Fig. 2b) are manifested in the increased variations in the water storage capacity profiles of the deeper Martian mantle (*wd*-dominated layers) in Fig. 4a–b.

The depth-dependent profiles (Fig. 4a–c) and $P$–$T$ water storage capacity diagram (Fig. 3a) show that the Martian mantle water storage capacity varies with depth, but a depth-dependent profile in water storage capacity does not necessarily imply a heterogeneous distribution of actual water content with depth. Instead, the water storage capacity in the solid mantle only indicates the maximum water content as a saturation limit for dehydration melting (Bercovici and Karato, 2003), which is not necessarily equivalent to the actual mantle water content. It has been suggested that the water distribution in the Earth's mantle is heterogeneous due to the presence of advection, such as partial melting upon transport to a region with a lower water storage capacity (Karato et al., 2020). Due to the lack of a sharp water storage capacity discontinuity in the Martian mantle, however, the associated partial melting mechanism would be less effective at maintaining a heterogeneous water distribution in Mars.

A recent Martian hydrogen isotope study (Barnes et al., 2020) characterized the D/H ratios of the Martian crust between 3.9 and 1.5 Ga based on ALH 84001 and NWA 7034 (D/H = 2.68–5.73 × $10^{-4}$). That study argued that to produce such D/H ratios for the crust using mixing models, at least two isotopically-distinct mantle water reservoirs with different water contents are required, implying that the Martian mantle has almost always been heterogeneous. If so, we cannot use Martian meteorite samples to infer the bulk mantle water content, because they may only represent the source regions where their parental magmas were generated (e.g., olivine-phyric shergottites were likely generated at 3–5 GPa or 255–425 km; Filiberto and Dasgupta, 2011). However, we may directly compare the water content in the mantle source regions of the Martian meteorites (15–23 ppm wt in the depleted shergottite source and 36–72 ppm wt in the enriched shergottite source; McCubbin et al., 2016) with the mantle water storage capacity at the same depth (480–650 ppm wt at $T_p$ = 1500–1900 K) to infer the hydration state of the source region (expressed as a percentage of the bulk water storage capacity at this depth), which yields 2–5% and 6–15% hydration for the depleted and enriched sources, respectively.

### 3.2 Estimates of the solid Martian mantle $H_2O$ storage capacity as a function of mantle $T_p$

The total water storage capacity integrated over depth for the solid Martian mantle was calculated for each $T_p$ from 1500 K to 1900 K (Fig. 4d–e and Table S6). For the present-day Martian mantle with $T_p$ = 1600 K, the bulk mantle water storage capacity is $9.0^{+2.8}_{-2.2}$ km GEL. If the Martian mantle was 100–200 K hotter at the beginning of the Noachian (~4.1 Ga; Filiberto, 2017), its mantle water storage capacity would have been $7.3^{+2.1}_{-1.9}$ km GEL and $6.1^{+1.9}_{-1.7}$ km GEL for $T_p$ =1700 K and 1800 K, respectively. These water storage capacities for the early Noachian ($T_p$ = 1700–1800 K) are approximately 1.7–2.9 km GEL smaller than the present-day water storage capacity ($T_p$ = 1600 K). The water storage capacity for the initial Martian mantle with $T_p$ = 1900 K can also be estimated to be $4.9^{+1.7}_{-1.5}$ km GEL.

To date, the scarcity of Martian rock samples has not allowed for a robust reconstruction of the planet's thermal history. Based on orbital spectroscopy of the Martian surface, the estimated mantle $T_p$ associated with its major volcanic provinces is ~1600–1650 K during the Amazonian (~3–0 Ga) and ~1650–1700 K during the Hesperian (~3.7–3 Ga) (Baratoux et al., 2011). Geochemical analyses of Noachian basalts in Gusev Crater, Gale Crater, and Meridiani Planum (~4.2–3.7 Ga), as well as clasts in the Martian meteorite NWA 7034 (pre-Noachian, ~4.4 Ga), suggest a range of average early mantle $T_p$ of 1600–1800 K (Filiberto and Dasgupta, 2015; Filiberto, 2017). Some *ol*-phyric shergottites indicate a much higher mantle $T_p$ of 1800–2000 K, but these samples might only represent a regional temperature anomaly (Filiberto, 2017). Therefore, we consider $T_p$ = 1600 K and 1800 K to be pertinent to the average Martian mantle today and at the beginning of the Noachian period (~4.1 Ga), respectively, with an estimated uncertainty of ±100 K. $T_p$ = 1900 K is prescribed as the highest $T_p$ corresponding to a mostly solidified Martian mantle, because the adiabatic areotherms at $T_p$ > 1900 K are significantly above the nominally anhydrous Martian mantle solidus (see Fig. S1 and references therein) and hence would cause extensive melting of the Martian upper mantle.

We adopted $T_p$ as a simple proxy for the Martian mantle thermal state and took the adiabats as a first-order approximation to the temperature gradient in the Martian mantle. This assumption is likely unrealistic for early Mars because its temperature gradients may have deviated from adiabats in the solid shallow mantle where they exceeded the solidus, causing partial melting (Fig. S1 and Rey (2015)). From the depths where the adiabats exceed the solidus to the base of the crust (~50 km), therefore, we only include the unmelted portion of the shallow mantle in our bulk mantle water storage capacity calculation. Consideration of a moderate temperature deviation of ~100 K (e.g., between the solid mantle adiabats and the melt adiabats, Fig. S1) in the shallow mantle contributes a small error to the water storage capacity of <9 m GEL, which is negligible compared to the mantle $H_2O$ storage capacities. However, such a simple shift from the adiabats may not be sufficient to reflect the actual lithospheric temperature deviation through time; conductive cooling from the surface leads to the formation of a lithospheric lid that thickens with time. The thickening of a much cooler lithospheric lid may result in a considerable temperature deviation of a few hundred kelvins and hence contribute a moderate error of tens of meters GEL to our bulk mantle water storage capacity estimates (at least 1–2 order of magnitude smaller than ~9 km GEL at $T_p$ = 1600 K). If water-saturated (with several wt% $H_2O$), a number of dense hydrous magnesium silicates (DHMSs) would stabilize at Martian mantle pressures (Ganskow and Langenhorst, 2014), especially within a much cooler lithospheric mantle (Wade et al., 2019). However, without efficient recycling of water back to the interior, Martian mantle assemblages would be water-undersaturated and primarily composed of NAMs only. Therefore, a thicker and cooler lithosphere does not substantially change our results.

In this study, we used the canonical bulk silicate Mars (BSM) model from Taylor (2013) (T13), but we do not mean to imply that the BSM composition is well-constrained at present. The differences between some recently-published BSM models (e.g., Taylor, 2013; Liebske and Khan, 2019; Yoshizaki and McDonough, 2020; Khan et al., 2022) are another potential source of uncertainty in our calculation. Two main parameters of the BSM models that affect their bulk mantle water storage capacities are Mg# (molar) and Mg/Si (in wt%) (Fig. 5 and Tables S6–S10). For example, with similar Mg#, the BSM model derived from mixing of meteoritic isotopic compositions (e.g., "EH+L" from LK19) has a lower Mg/Si ratio than T13, resulting in less olivine and more pyroxene in the Martian mantle (Fig. S3), which lowers the bulk mantle water storage capacity by 0.4 km GEL (<5%) at $T_p$ = 1600 K (Fig. 5). Alternatively, with the same Mg/Si, YM20 proposed an FeO-poor BSM model with a higher bulk silicate Mg# of 79, in contrast to the T13 model, with a bulk silicate Mg# of 75. As discussed in Section 3.1, the effects of Fe on *ol* and *wd* are typically positive, and the effect of iron on *rw* is negligible or slightly negative, so the bulk mantle water storage capacity based on YM20 is lower than that based on T13 by 1.3 km GEL (<15%) at $T_p$ = 1600 K (Fig. 5). The five different BSM models we selected represent the full possible range of bulk silicate Mg# and Mg/Si. We tested the sensitivity of our bulk mantle water storage capacity estimates to BSM composition at $T_p$ = 1600 K and 1900 K (Fig. 5, S8 and Tables S6–S10) and found the effects of mantle composition contribute an error of <5–15% (Fig. 5 and S8). A detailed discussion of the effects of bulk composition on Martian mantle mineralogy and areotherms is beyond the scope of this study.

### 3.3 Comparing the solid mantle $H_2O$ storage capacities of Mars and Earth and their evolution through time

For Earth's mantle with a bulk silicate Mg# of ~89, the water storage capacity of the *ol*-dominated upper mantle remains ~4–6 orders of magnitude lower than that of the *wd*-dominated transition zone at similar *P–T* conditions where *ol* and *wd* coexist (~13–15 GPa), forming a sharp water storage capacity increase at the 410-km discontinuity of Earth's mantle (Fig. 3b; Dong et al., 2021). In contrast, the *ol*-dominated upper portion of the Martian mantle (bulk silicate Mg# ≈ 75) (Fig. 3a) can incorporate much more water than Earth's upper mantle (bulk silicate Mg# ≈ 89) due to the positive effect of Fe on olivine water storage capacity. The water storage capacity of the Martian mantle *ol*-dominated layer can reach and even exceed that of the *wd*-dominated layer (Fig. 4a–c). Even though the increased abundance of Fe in *rw* does not change its water storage capacity significantly, the bulk water storage capacity in the lowermost Martian mantle (*rw*-dominated layer) can be much higher than that in the Earth at similar pressures when the temperature is high. This is because the equilibrium phase assemblage of the Earth's mantle is mostly comprised of *gt* and ferropericlase (*fp*) above 2100 K at 20 GPa (Fig. 3b), and *gt* + *fp* has a much lower water storage capacity than *rw*. In contrast, the high Fe content of the Martian mantle stabilizes *rw* up to the solidus, with no *gt* + *fp* field present in the equilibrium phase diagram (Fig. 3a).

The bulk water storage capacity of the Martian mantle increases with decreasing $T_p$ (Fig. 4), as in the Earth (Dong et al., 2021), which mainly results from the negative effects of temperature on the water storage capacities of *ol*, *wd*, and *rw*. The evolution of bulk water storage capacity in Mars and Earth as a function of $T_p$ during secular cooling are compared in Fig. 4e. On the one hand, with $T_p = 1600$ K (present-day), the bulk water storage capacity of the Earth's mantle (~2.3 OM, where OM is the modern surface ocean mass on Earth, $1.335 \times 10^{21}$ kg $H_2O$) is only a factor of ~2.3 larger than that of Mars (~1.0 OM), despite the volume of Earth's mantle (~9.1×10$^{11}$ km$^3$) being nearly ~6.5 times larger than that of Mars (~1.4×10$^{11}$ km$^3$). This difference arises primarily from the Fe enrichment in the Martian mantle. On the other hand, the bulk water storage capacities of Earth and Mars converge at higher $T_p$ (Fig. 4e, 0.4–0.6 Earth's ocean masses at $T_p = $ ~1950 K), because 1) the effects of temperature on the water storage capacities of NAMs outcompete the effects of iron at high $T_p$, and 2) contributions of Earth's lower mantle to its bulk mantle water storage diminish significantly at high $T_p$. Last but not least, though the evolution of mantle water storage capacities in Earth and Mars follows the same increasing trend with secular cooling (Fig. 4e), mantle rehydration by subduction (e.g., van Keken et al., 2011) may have been increasing Earth's actual mantle water content since the onset of plate tectonics (light blue dashed arrow in Fig. 4e). In contrast, Mars lacks efficient water recycling mechanisms such as long-term subduction-driven plate tectonics (Nimmo and Tanaka, 2005), so its actual mantle water content may have remained relatively low to the present (light red dashed arrow in Fig. 4e).

### 3.4 Possible sources and sinks for water in the Martian mantle through time

During accretion, Mars acquired its primordial $H_2O/H_2$ from chondritic materials (Brasser, 2013) and/or the solar nebula (Olson and Sharp, 2018). Some fraction of this primordial water was retained in the Martian mantle upon solidification of the magma ocean (e.g., Elkins-Tanton, 2008; Kurokawa et al., 2021). Subsequently, volcanic outgassing continued to release water onto the surface (e.g., Craddock and Greeley, 2009; Phillips et al., 2001). The initial water content in the newly-solidified mantle was limited by its bulk water storage capacity of ~5 km GEL at $T_p =$ 1900 K. Assuming that the early Martian hydrosphere was supplied by outgassing only, the amount of water estimated for the putative early oceans (~0.1–2.4 km GEL) (e.g., Carr and Head, 2003; Sholes et al., 2021; Fig. 6) requires the initial Martian mantle reservoir to outgas > ~2–48% of its water, and to fill up all global valley networks of ~0.6–1.8 km GEL (e.g., Rosenberg et al., 2019; Luo et al., 2019; Fig. 6) would require the initial Martian mantle reservoir to outgas > ~12–36% of its water.

The early hydrosphere initially outgassed from the mantle was lost either to space through atmospheric escape (e.g., Kurokawa et al., 2014) or to the crust through serpentinization (e.g., Chassefière et al. 2013), or, most likely, to both (e.g., Scheller et al., 2021). A return flux of water into the crust is estimated to be ~0.1–1.5 km GEL $H_2O$ based on Martian surface D/H (Chassefière et al., 2013; Scheller et al., 2021), which is comparable either to the storage capacity of ~0.2–0.9 km GEL $H_2O$ in the top 5–10 km of hydrated minerals in the Martian crust (e.g., Mustard et al., 2019; Wernicke and Jakosky, 2021; Fig. 6) or to the storage capacity of ~0.1–1.0 km GEL $H_2O$ in the available pore space of the permeable Martian crust to a depth of ~26.5 km (e.g., Clifford and Parker, 2001; Clifford et al., 2010; Fig. 6). Such a return flux would be insufficient to hydrate the lower crust and hence return water back to rehydrate the mantle, despite the relatively large water storage capacity estimated for Martian basaltic materials at shallow mantle conditions (Wade et al., 2017).

O'Rourke et al. (2019) proposed that water stored in the *rw*-dominant layer at the base of the Martian mantle may react with and enter the core as $FeH_x$. However, the nominally anhydrous *rw* layer in direct contact with the Fe-rich core would soon become enriched in FeO and convert into a layer of dry iron-rich *fp* and stishovite (*st*) if the core was reacting with the base of the mantle (Matsuzaka et al., 2000), which would impede any further hydrogenation of the core. Therefore, a significant exchange of water between the Martian mantle and core since their formation remains unlikely.

## 4 Summary and conclusions

In this study, we applied a bootstrap aggregation method to model the effects of Fe on the water storage capacities of major terrestrial mantle NAMs, *ol*, *wd*, and *rw*, based on high-pressure mineral physics data in the literature. We assessed the variations of our parameterization of the Fe effects based on the errors in the experimental data on water measurements in NAMs. We found that our bootstrapped resampling typically indicates that iron increases the water storage capacities of *ol* and *wd* and has no/little effect on the water storage capacity of *rw*.

We presented depth-dependent water storage capacity profiles for the Martian mantle and found that the water storage capacity of the more Fe-rich Martian mantle is in general higher than that of Earth at similar *P*–*T*, due to iron increasing the water storage capacities of *ol* and *wd* as well as expanding the *rw* stability field to higher temperatures. We also calculated the bulk water storage capacity of the solid Martian mantle as a function of mantle $T_p$. As the mantle cooled, its water storage capacity would have increased by 4.1 km GEL, from 4.9 km GEL ($T_p$ = 1900 K, initial mantle) to 9.0 km GEL ($T_p$ = 1600 K, present-day). We tested the sensitivity of those estimates to BSM composition and found the effects of mantle composition contribute an error of <5–15%. Despite an increase in its water storage capacity through time, the actual water content in the Martian mantle today is lower than what it was in the pre-Noachian with a net water loss to the surface, due to lack of an efficient deep-water recycling mechanism on Mars.

**Data Statement**

The datasets on water storage capacities in NAMs used in these calculations are compiled from the literature, and are available within the Zenodo repository (https://doi.org/10.5281/zenodo.4976454) or Table S11–S13 in the supporting information. The thermodynamic code HeFESTo used in these calculations is available on GitHub (https://github.com/stixrude/HeFESToRepository) or within the Zenodo repository (https://doi.org/10.5281/zenodo.5014204).

**Acknowledgments**

We thank the editor, Elizabeth Rampe, for handling our manuscript, and we thank the reviewers, Thomas Ruedas and J. Brian Balta, for constructive comments and the former for his generous checking of our equations and text; responsibility for any remaining errors rests solely with the authors. We also thank Robin Wordsworth for helpful advice and Sam Klug (The Fellowships & Writing Center at Harvard) for improving the writing of this manuscript. This work is derived from a final project prepared by J. Dong and Z.T. Eriksen for a Harvard seminar course taught by R.A. Fischer in Fall 2018 (E-PSCI 248: Topics in Mineral Physics and Chemistry: Volatiles in the Deep Earth), and it benefited from the seminar discussions. The preliminary results of this work were presented at the 2019 Goldschmidt conference with the support of a planetary science travel grant from the National Aeronautics and Space Administration. J. Dong was supported by a James Mills Peirce Fellowship from the Graduate School of Arts and Sciences at Harvard University. R.A. Fischer was supported by the Henry Luce Foundation. L.P. Stixrude was supported by the European Research Council under Advanced Grant 291432 MoltenEarth (FP7) and by the National Science Foundation (NSF) (EAR-1853388). C.R. Lithgow-Bertelloni was supported by research funding from the Louis B. and Martha B. Slichter Endowed Chair in the Geosciences at University of California Los Angeles and by the NSF (EAR-1900633). M.C. Brennan was supported by an NSF Graduate Research Fellowship (DGE-1745303).

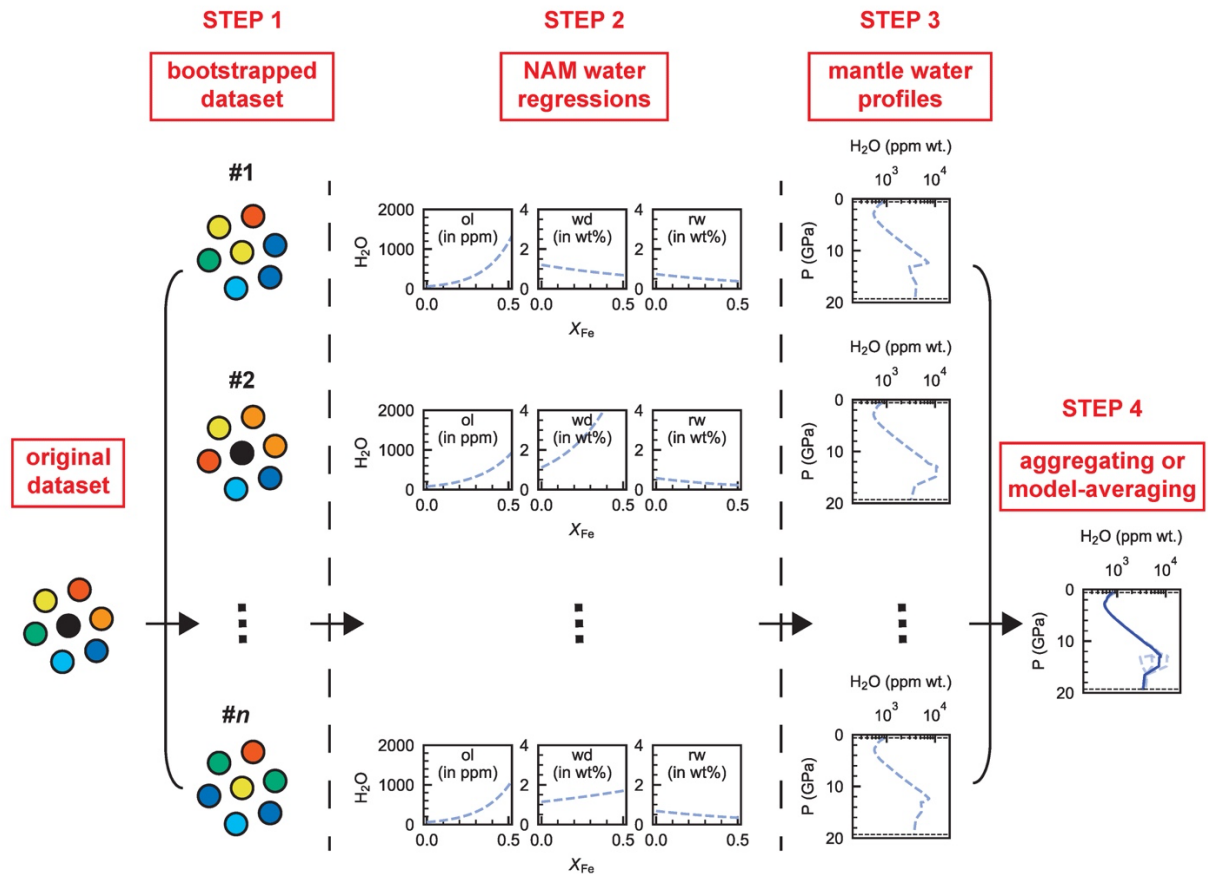

**Figure 1**: Workflow for the "bagging" method. Step 1: resampling from the original datasets for *ol*, *wd*, and *rw* (compiled in this study) with replacement. Step 2: fitting the *T*, *P*, and $X_{Fe}$ dependence for each mineral for each of *n* resampled datasets. Step 3: calculating *n* bulk water storage capacity profiles based on varying *T*, *P*, $X_{Fe}$, and phase proportions along a specific areotherm at each $T_p$. Step 4: averaging all *n* profiles to obtain the best-fit bulk water storage capacity profile for $T_p$. NAMs are labeled as: *ol*→olivine, *wd*→wadsleyite, *rw*→ringwoodite.

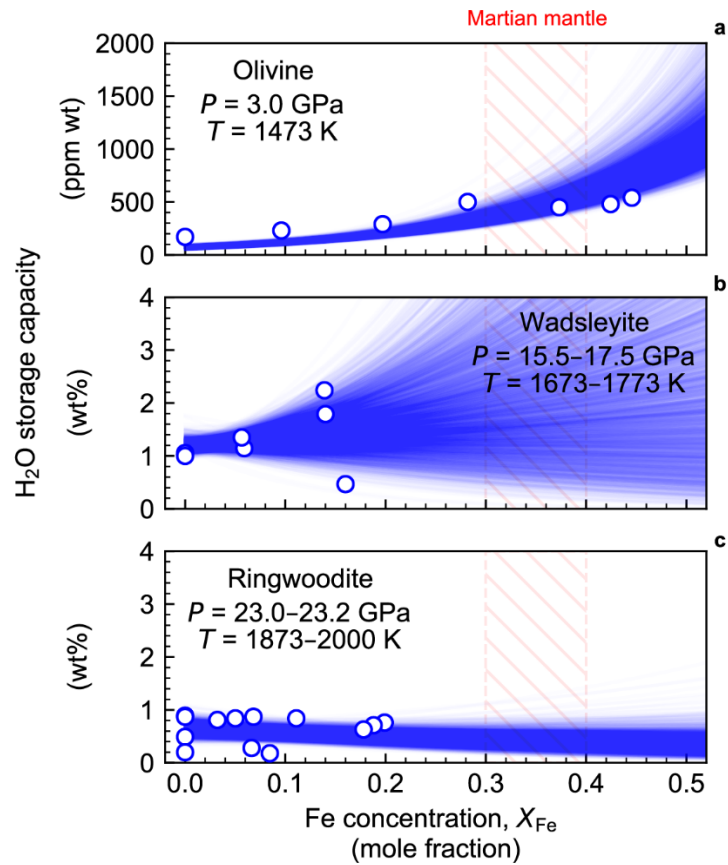

**Figure 2**: Effects of Fe content on the water storage capacities of (a) olivine, (b) wadsleyite, and (c) ringwoodite at selected pressures and temperatures. The open circles are experimentally-determined water storage capacity data from the narrow range of $P$–$T$ conditions shown in the text of each figure for comparison to our models (Table S11–S13 and references therein); their associated uncertainties are not shown for clarity. The blue curves correspond to the water storage capacities predicted at the same $P$–$T$ conditions for each mineral from the regressions of $10^4$ sets of bootstrapped data (Section 2.3). The red hatched regions indicate the approximate range of $X_{Fe}$ expected for olivine, wadsleyite, and ringwoodite in the solid Martian mantle.

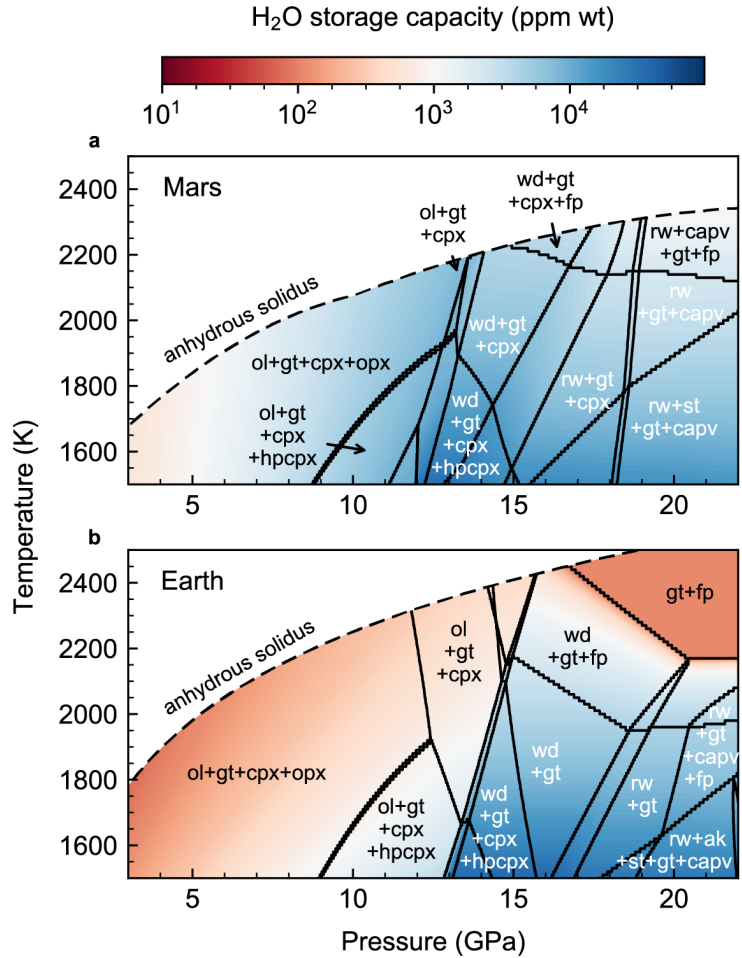

**Figure 3**: H$_2$O storage capacity diagrams of Martian (a; model-averaged estimate from this study) and terrestrial (b; Dong et al., 2021) mantle assemblages between 3 and 22 GPa. The bulk H$_2$O storage capacities at each $P$–$T$ were estimated from the H$_2$O storage capacities of the stable phases and their relative abundances. The thermodynamically-stable mantle assemblages were calculated for a bulk silicate Mars composition (Taylor, 2013) and a terrestrial depleted MORB mantle composition (Workman and Hart, 2005) using the thermodynamic code HeFESTo (Stixrude and Lithgow-Bertelloni, 2005, 2011). The anhydrous mantle solidi for Mars and Earth are from Duncan et al. (2018) and Herzberg et al. (2000), respectively. Mantle phases are labeled as: *ol*→olivine, *wd*→wadsleyite, *rw*→ringwoodite, *fp*→ferropericlase, *cpx*→clinopyroxene, *opx*→orthopyroxene, *hpcpx*→high-pressure clinopyroxene, *ak*→akimotoite, *st*→stishovite, and *capv*→CaSiO$_3$ perovskite.

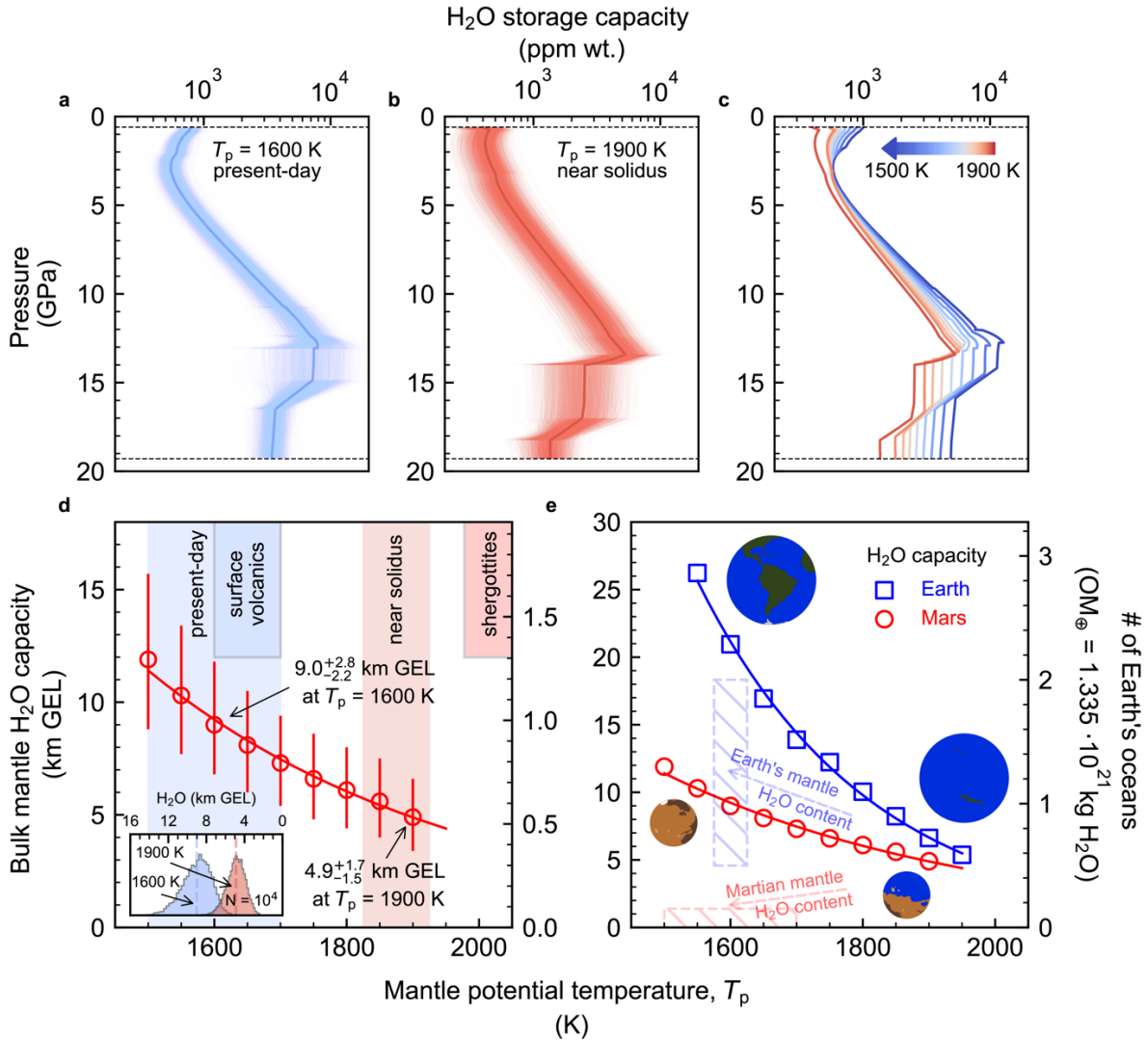

**Figure 4**: Water storage capacity of the solid Martian mantle. In the upper panels, the bold lines are the best-fit water storage capacity profiles along mantle adiabats at (a) $T_p$ = 1600 K, (b) $T_p$ = 1900 K, and (c) $T_p$ = 1500–1900 K, in increments of 50 K. The light curves in (a) and (b) are $10^4$ profiles from a representative set of "bagging" sampling. In the lower panels, the red open circles are the bulk water storage capacities for the solid Martian mantle at each $T_p$, which were calculated as the medians of $10^4$ "bagging" samples; the blue open squares are the bulk water storage capacities for the solid Earth's mantle estimated by Dong et al. (2021). In (d), the error bars represent the 5th and 95th percentiles of each distribution; the shaded areas are the estimated ranges for the Martian mantle $T_p$ (Table S6); and the boxes are the $T_p$ ranges estimated for different Martian basalt suites (Filiberto, 2017). In the inset of (d), two distributions of "bagging" samples are shown, for $T_p$ = 1600 K (blue) and $T_p$ = 1900 K (red). The vertical dashed lines are the best-fit (model average) bulk water storage capacity for each $T_p$. In (e), the hatched rectangular regions correspond to geochemical estimates of the actual

mantle water contents of Mars (red, 14–250 ppm wt or 80–1460 m GEL; Filiberto, 2019) and Earth (blue, 0.5–2 OM or 5–19 km GEL; Dauphas and Morbidelli, 2014; Hirschmann, 2018).

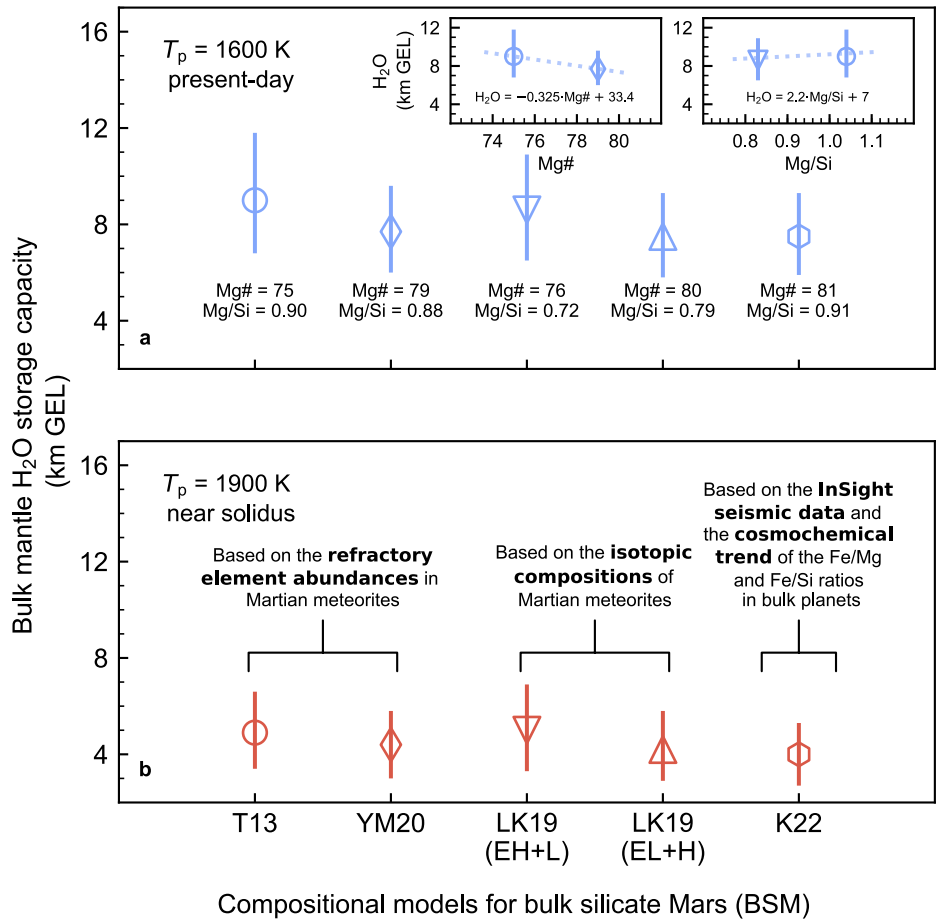

**Figure 5:** Sensitivity of bulk mantle $H_2O$ storage capacity to mantle composition at (a) 1600 K (blue) and (b) 1900 K (red). T13 and LK19 (EL+H) with an Mg# of 75–76 are more enriched in FeO than YM20, LK19 (EH+L), and K22 with an Mg# of 79–81. T13, YM20, and K22 have an Mg/Si of 1.04–1.05, similar to Earth. In contrast, two LK19 compositions have a lower Mg/Si of 0.83 and 0.91, leading to lower olivine and higher pyroxene abundances in the calculated Martian mantle assemblages (Fig. S2–S6). The canonical T13 composition and the other four compositions represent the entire possible range of the bulk Mg# and Mg/Si for the Martian mantle. Within this range, the bulk $H_2O$ storage capacity of the Martian mantle decreases slightly with increasing bulk Mg# and increases slightly with increasing Mg/Si (insets, a), but in general it varies by less than ~13–17% (Table S6–10). Hence, our estimates based on T13 in Fig. 4 are reasonable first-order approximations.

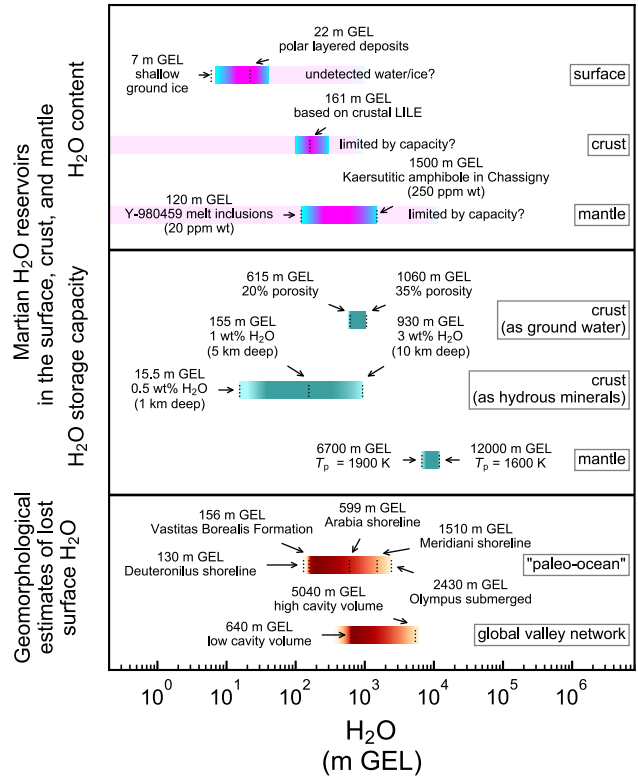

**Figure 6:** Summary of the Martian water reservoirs in the surface, crust, and mantle (water content in blue/pink and water storage capacity in blue-green), as well as the estimates of the lost surface water potentially associated with morphological features on Mars (red/orange). The gradients in color represent the qualitative uncertainty for each bar. In general, these estimates are subject to considerable uncertainty, but their orders of magnitude are much more robust. Estimates from: Carr and Head (2015) (polar layered deposits and shallow ground ice), Barnes et al. (2020) (crustal $H_2O$ content based on enriched shergottite mantle and large ion lithophile elements (LILE) crustal abundance), Filiberto et al. (2019) (mantle source region $H_2O$ content based on Yamato 980459 melt inclusions and Kaersutitic amphibole in Chassigny), Clifford et al. (2010) (pore water volumes of the crust for different porosities), Mustard et al. (2019) (hydrated minerals for different depths of crust), this study (bulk mantle water storage capacity), Carr (2003) (volumes of ocean contained by putative shorelines for different elevations), Rosenberg et al. (2019) and Luo et al. (2020) (volume of water required to carve the Martian valley networks for different cavities).

Supporting Information for

**Water storage capacity of the Martian mantle through time**

Junjie Dong[1][2]*, Rebecca A. Fischer[1], Lars P. Stixrude[3], Carolina R. Lithgow-Bertelloni[3], Zachary T. Eriksen[1], Matthew C. Brennan[1]

[1] Department of Earth and Planetary Sciences, Harvard University, Cambridge, Massachusetts 02138, USA.
[2] Department of the History of Science, Harvard University, Cambridge, Massachusetts 02138, USA.
[3] Department of Earth, Planetary, and Space Sciences, University of California, Los Angeles, California 90095, USA.

**Contents of this file**

**Supplementary Text**

**S1. Density–pressure profiles for the Martian mantle**

Conversion from pressure to depth was performed using a thin shell calculation based on the density–pressure profiles pertinent to the potential temperature and bulk composition:

$$h_{i+1} = \frac{(P_{i+1} - P_i)}{\frac{(\rho_{i+1} + \rho_i)}{2} g_i} = r_i - r_{i+1} \quad (1)$$

$$V_{i+1} = \frac{4\pi}{3} \left[ r_i^3 - (r_i - h_{i+1})^3 \right] = \frac{4\pi}{3} (r_i^3 - r_{i+1}^3) \quad (2)$$

$$g_{i+1} = \frac{G\left(M - \sum_{i=0}^{n} V_{n+1} \frac{\rho_{n+1} + \rho_n}{2}\right)}{(r_i - h_{i+1})^2} = \frac{G\left(M - \sum_{i=0}^{n} V_{n+1} \frac{\rho_{n+1} + \rho_n}{2}\right)}{r_{i+1}^2} \quad (3)$$

where $G$ is the gravitational constant ($G = 6.67408 \times 10^{-11}$ m$^3$/kg/s$^2$); the mass $M$ and radius $r_0$ of Mars are $6.4171 \times 10^{23}$ kg (Konopliv et al., 2011) and 3390 km, respectively; and the fictitious density of Martian surface rock $\rho_0$ is 3 kg/m$^3$.

## Supplementary Figures

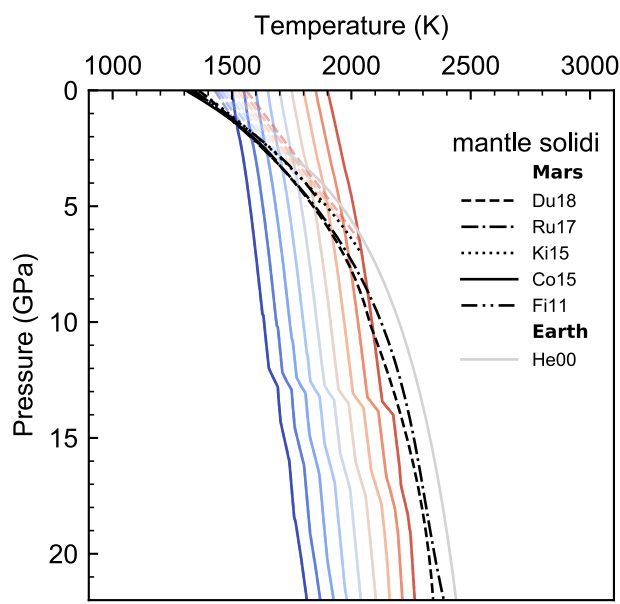

**Figure S1:** Adiabatic temperature gradients (colored solid lines) of the "**T13**" Martian mantle computed with HeFESTo (Stixrude and Lithgow-Bertelloni, 2005, 2011). Adiabats are shown for potential temperatures of 1500–1900 K, with a spacing of 50 K. When adiabats of the convective solid mantle intersect its solidus, the actual mantle temperature will either continue to follow the adiabats if the melts remain within the source rocks (e.g., in a deep mantle upwelling), or the temperature will be deflected to follow the mantle solidus (e.g., in the case of lithospheric thinning) as the primary melts produced at the solidus temperature are extracted and ascend along their respective adiabats (colored dashed lines; Collinet, 2021). Which of these two cases will arise depends on the geodynamic setting of mantle melting (cf. Rey et al., 2015). The geodynamics of Martian mantle melting remain poorly constrained, and in this study the solid mantle adiabats are taken as a first-order approximation to the temperature gradients of the shallow mantle where the adiabats exceed the solidus. A temperature variation of ~100 K (e.g., between the solid mantle adiabats and the melt adiabats) in the shallow mantle contributes an uncertainty of <0.1% of one Earth's ocean mass (<0.001 OM) and is negligible. Only the unmelted portion of the shallow mantle is included in the bulk mantle water storage capacity calculations. The respective melt fraction at a specific $P$–$T$ is estimated based on Collinet et al. (2015). Several Martian mantle solidi from the literature are shown as black curves: Duncan et al. (2018) (dashed), Ruedas and Breuer (2017) (dashdotted), Kiefer et al. (2015) (dotted), Collinet et al. (2015) (solid), and Filiberto and Dasgupta (2011) (dot-dot-dashed). The solid grey curve is the solidus for Earth's mantle from Herzberg et al. (2000).

## <u>T13</u>

**a**

### $T_p$ = 1600 K

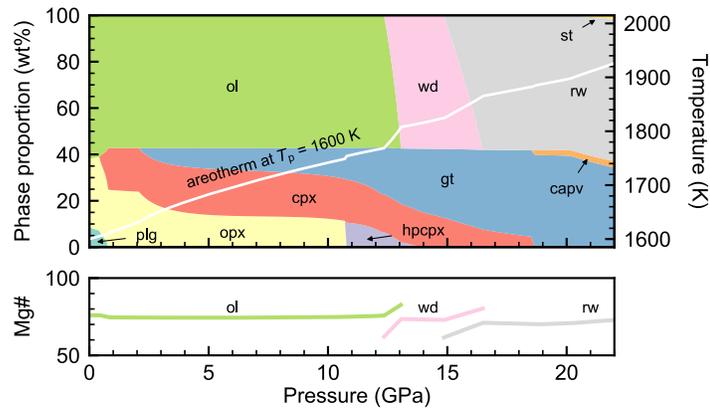

**b**

### $T_p$ = 1900 K

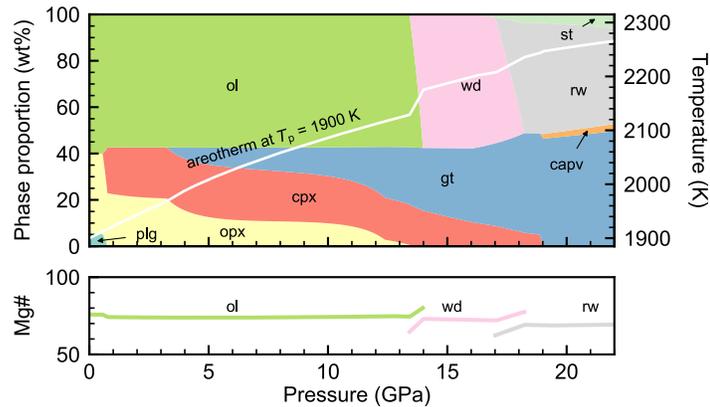

**Figure S2:** Examples of calculated phase assemblages for the "**T13**" Martian mantle along the 1600 K (a) and 1900 K (b) adiabats. The upper panels show modal phase proportions along the adiabats (colored regions, left axes) and the temperature along the adiabat (white curves, right axes). The lower panels show the Mg# of olivine, wadsleyite, and ringwoodite along the adiabats.

## LK19 (EH+L)

**a**

### $T_p$ = 1600 K

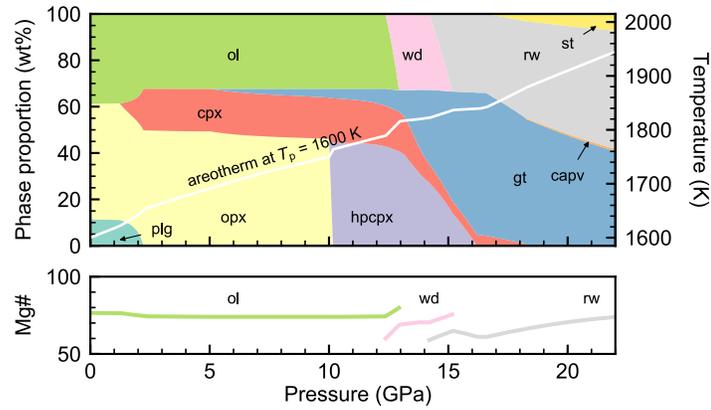

**b**

### $T_p$ = 1900 K

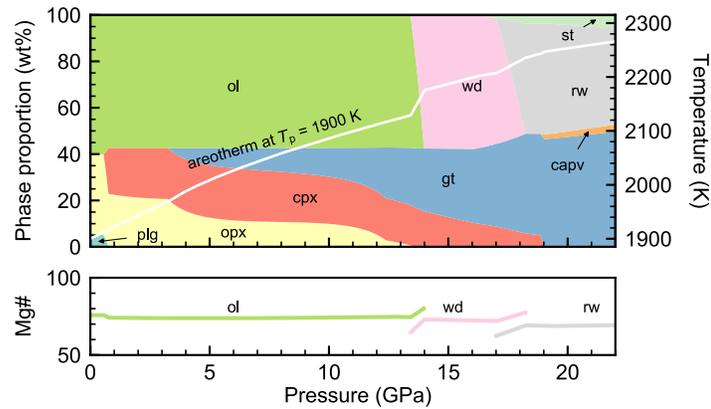

**Figure S3:** Examples of calculated phase assemblages for the "**LK19 (EH+L)**" Martian mantle along the 1600 K (a) and 1900 K (b) adiabats. The upper panels show modal phase proportions along the adiabats (colored regions, left axes) and the temperature along the adiabat (white curves, right axes). The lower panels show the Mg# of olivine, wadsleyite, and ringwoodite along the adiabats.

## **LK19 (EL+H)**

**a**

### $T_p$ = 1600 K

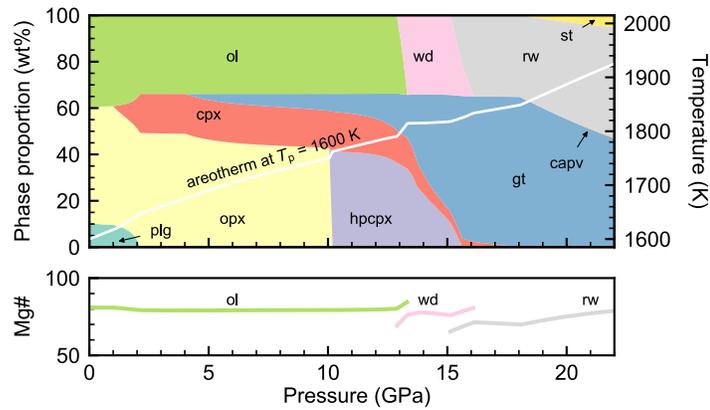

**b**

### $T_p$ = 1900 K

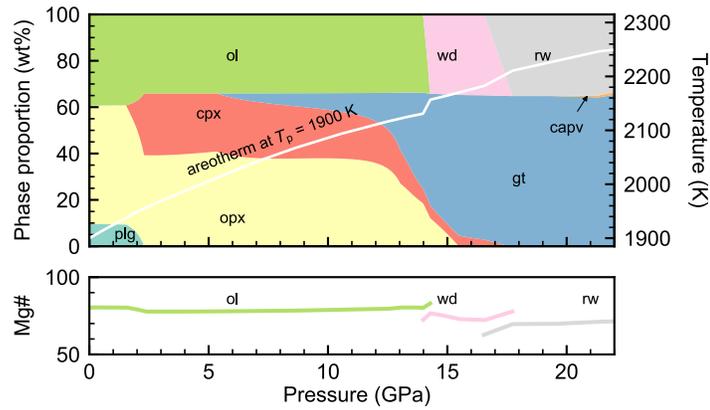

**Figure S4:** Examples of calculated phase assemblages for the "**LK19 (EL+H)**" Martian mantle along the 1600 K (a) and 1900 K (b) adiabats. The upper panels show modal phase proportions along the adiabats (colored regions, left axes) and the temperature along the adiabat (white curves, right axes). The lower panels show the Mg# of olivine, wadsleyite, and ringwoodite along the adiabats.

## **YM20**

**a**

### $T_p$ = 1600 K

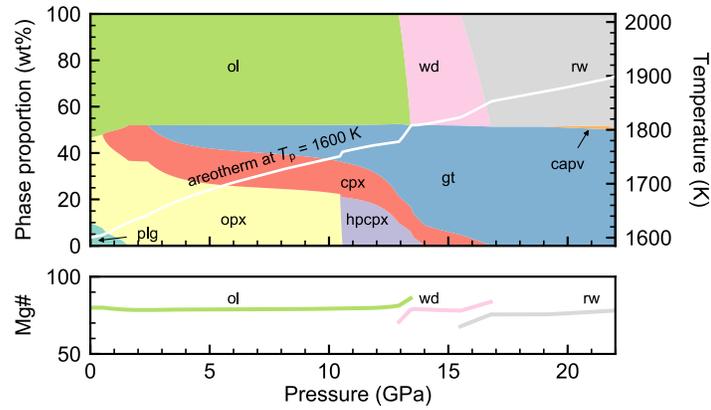

**b**

### $T_p$ = 1900 K

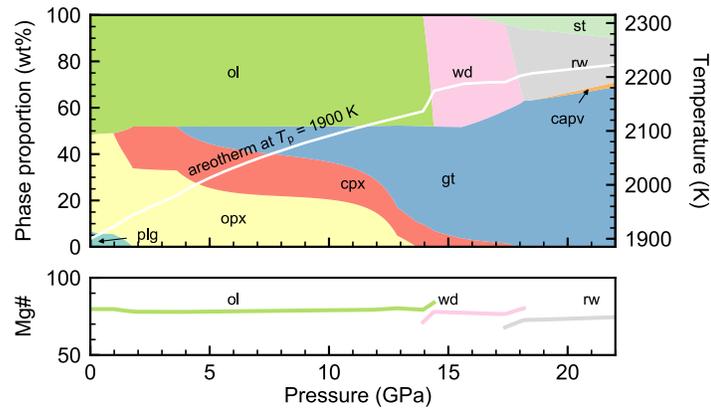

**Figure S5:** Examples of calculated phase assemblages for the "**YM20**" Martian mantle along the 1600 K (a) and 1900 K (b) adiabats. The upper panels show modal phase proportions along the adiabats (colored regions, left axes) and the temperature along the adiabat (white curves, right axes). The lower panels show the Mg# of olivine, wadsleyite, and ringwoodite along the adiabats.

## **K22**

**a**

$T_\text{p}$ = 1600 K

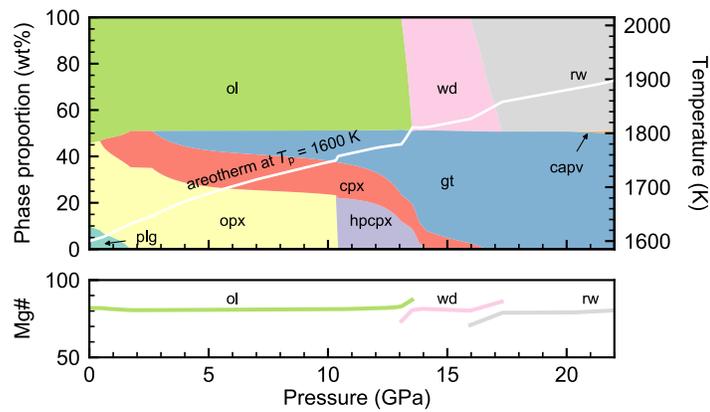

**b**

$T_\text{p}$ = 1900 K

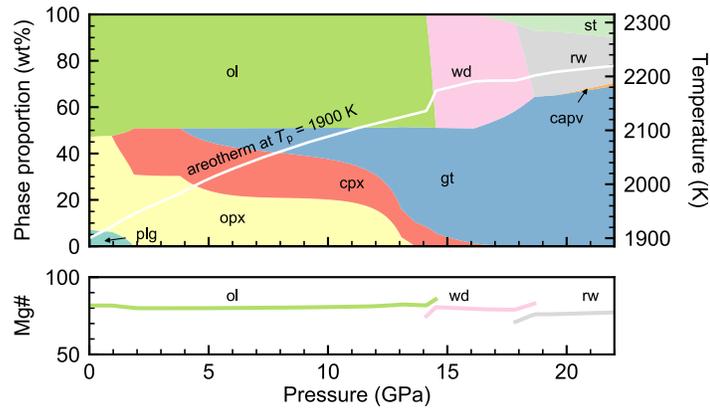

**Figure S6:** Examples of calculated phase assemblages for the "**K22**" Martian mantle along the 1600 K (a) and 1900 K (b) adiabats. The upper panels show modal phase proportions along the adiabats (colored regions, left axes) and the temperature along the adiabat (white curves, right axes). The lower panels show the Mg# of olivine, wadsleyite, and ringwoodite along the adiabats.

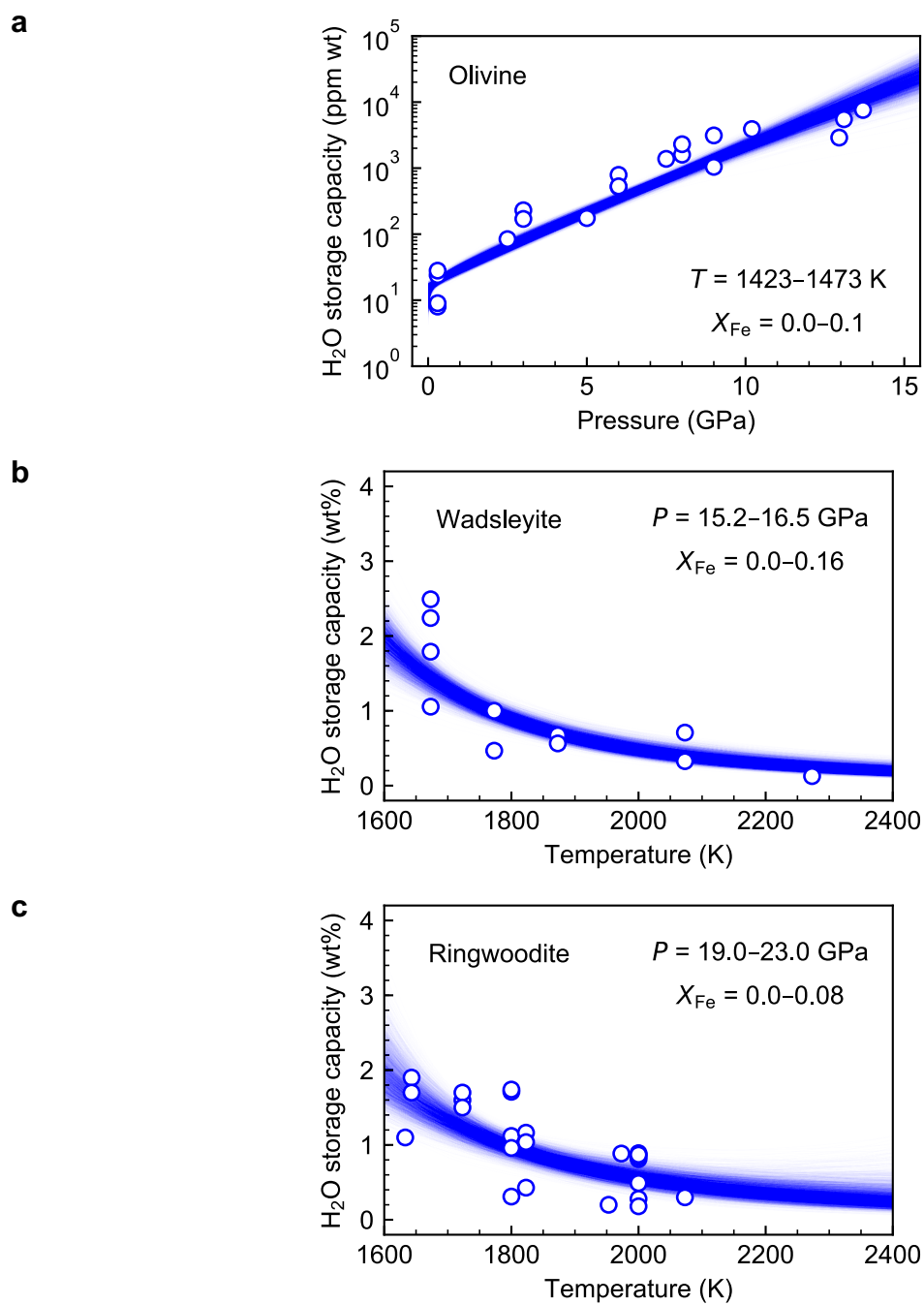

**Figure S7:** Pressure or temperature effects on the water storage capacities of (a) olivine, (b) wadsleyite, and (c) ringwoodite at selected *P–T* conditions. The open circles are a subset of the experimentally-determined water storage capacity data, at these *P–T*. Their associated error bars are not displayed for clarity. The blue curves correspond to $10^4$ regressions in a representative set of bootstrapping (see Section 2 of the main text for details).

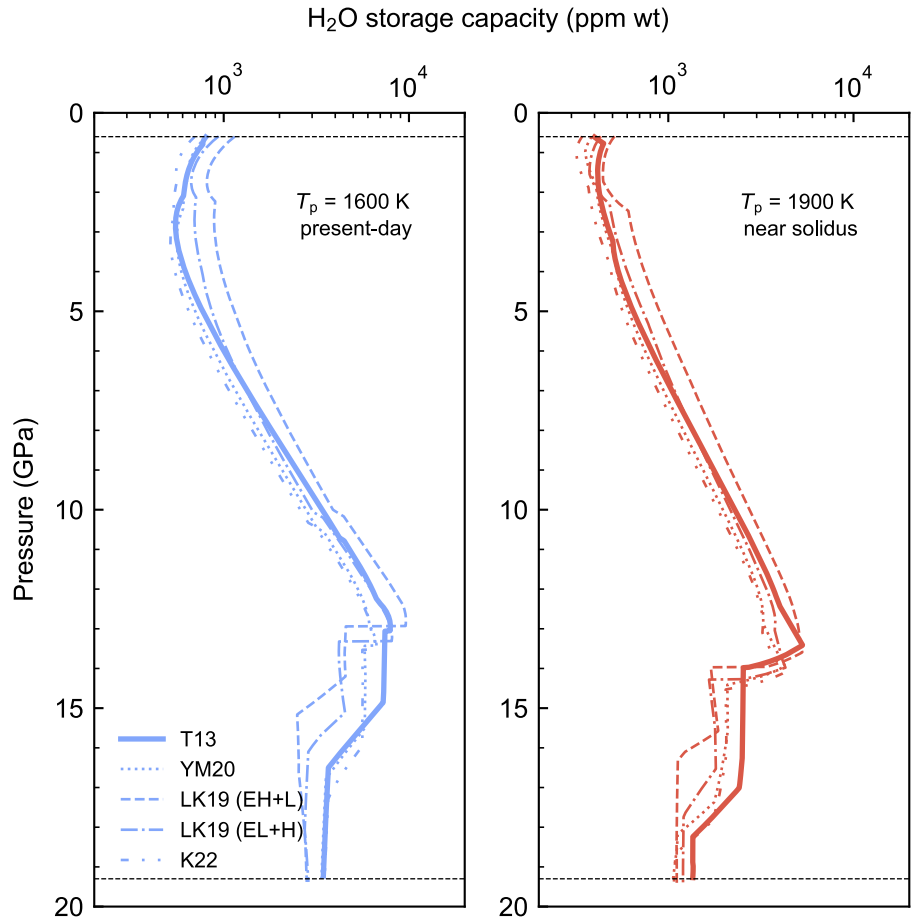

**Figure S8:** Best-fit water storage capacity profiles along mantle adiabats for five BSM compositions at (a) $T_p$ = 1600 K in blue and (b) $T_p$ = 1900 K in red. "T13": bold solid, "YM20": dotted, "LK19 (EH+L)": dashed, "LK19 (EL+H)": dot-dashed, and "K22": dot-dot-dashed.

**Table S1. Bulk Silicate Mars composition based on Taylor (2013)**

| component | oxides (wt%) | component | cations (10 atoms) |
|---|---|---|---|
| MgO | 30.5 | Mg | 4.07714 |
| $Al_2O_3$ | 3 | Al | 0.32128 |
| $SiO_2$ | 43.7 | Si | 3.91862 |
| CaO | 2.4 | Ca | 0.23347 |
| $TiO_2$ | 0.1 | Ti | - |
| FeO | 18.1 | Fe | 1.35735 |
| $Na_2O$ | 0.5 | Na | 0.09214 |
| $P_2O_5$ | 0.2 | P | - |
| $Cr_2O_3$ | 0.7 | Cr | - |
| MnO | 0.4 | Mn | - |
| Total | 99.8 | | 10 |
| | | | |
| **Mg# (molar)** | 75 | **Mg/Si (wt)** | 0.90 |

**Table S2. Bulk Silicate Mars composition based on Liebske and Khan (2019) ("EH+L")**

| component | oxides (wt%) | component | cations (10 atoms) |
|:---:|:---:|:---:|:---:|
| MgO | 27.9 | Mg | 3.68255 |
| $Al_2O_3$ | 2.8 | Al | 0.29218 |
| $SiO_2$ | 49.9 | Si | 4.41816 |
| CaO | 2.1 | Ca | 0.19922 |
| FeO | 16 | Fe | 1.18474 |
| $Na_2O$ | 1.3 | Na | 0.22316 |
| Total | 100 | | 10 |
| | | | |
| **Mg# (molar)** | 76 | **Mg/Si (wt)** | 0.72 |

**Table S3. Bulk Silicate Mars composition based on Liebske and Khan (2019) ("EL+H")**

| component | oxides (wt%) | component | cations (10 atoms) |
|---|---|---|---|
| MgO | 30.5 | Mg | 3.96081 |
| $Al_2O_3$ | 3.1 | Al | 0.31827 |
| $SiO_2$ | 50 | Si | 4.35563 |
| CaO | 2.1 | Ca | 0.19600 |
| FeO | 13.5 | Fe | 0.98350 |
| $Na_2O$ | 1.1 | Na | 0.18578 |
| Total | 100.3 | | 10 |
| | | | |
| **Mg# (molar)** | 80 | **Mg/Si (wt)** | 0.79 |

**Table S4. Bulk Silicate Mars composition based on Yoshizaki and McDonough (2020)**

| component | oxides (wt%) | component | cations (10 atoms) |
|-----------|--------------|-----------|--------------------|
| MgO | 31 | Mg | 4.06275 |
| $Al_2O_3$ | 3.59 | Al | 0.43313 |
| $SiO_2$ | 45.5 | Si | 4.02396 |
| CaO | 2.88 | Ca | 0.27128 |
| $TiO_2$ | 0.17 | Ti | - |
| FeO | 14.7 | Fe | 1.10832 |
| $Na_2O$ | 0.59 | Na | 0.10056 |
| $P_2O_5$ | 0.17 | P | - |
| $Cr_2O_3$ | 0.88 | Cr | - |
| MnO | 0.37 | Mn | - |
| Total | 99.85 | | 10 |
| | | | |
| **Mg# (molar)** | 79 | **Mg/Si (wt)** | 0.88 |

**Table S5. Bulk Silicate Mars composition based on Khan et al. (2022)**

| component | oxides (wt%) | component | cations (10 atoms) |
|---|---|---|---|
| MgO | 32.81 | Mg | 4.24162 |
| Al$_2$O$_3$ | 3.49 | Al | 0.35670 |
| SiO$_2$ | 46.66 | Si | 4.04638 |
| CaO | 2.66 | Ca | 0.24716 |
| FeO | 13.68 | Fe | 0.99213 |
| Na$_2$O | 0.69 | Na | 0.11601 |
| Total | 99.99 | | 10 |
| | | | |
| **Mg# (molar)** | 81 | **Mg/Si (wt)** | 0.91 |

**Table S6. Bulk H$_2$O storage capacity in the solid Martian mantle with the "T13" composition (km GEL\*)**

| $T_p$ | best-fit | 1$^{st}$ | 5$^{th}$ | 10$^{th}$ | 50$^{th}$ percentile | 90$^{th}$ | 95$^{th}$ | 99$^{th}$ |
|---|---|---|---|---|---|---|---|---|
| (K) | (mean) | | | | (median) | | | |
| 1500 | 11.9 | 7.9 | 8.8 | 9.3 | 11.6 | 14.8 | 15.7 | 17.7 |
| 1550 | 10.3 | 6.9 | 7.7 | 8.1 | 10.1 | 12.7 | 13.4 | 15.1 |
| 1600 | 9.0 | 6.0 | 6.8 | 7.2 | 8.9 | 11.1 | 11.8 | 13.1 |
| 1650 | 8.1 | 5.3 | 6.0 | 6.4 | 7.9 | 9.8 | 10.5 | 11.7 |
| 1700 | 7.3 | 4.7 | 5.4 | 5.7 | 7.2 | 8.9 | 9.4 | 10.6 |
| 1750 | 6.6 | 4.2 | 4.8 | 5.2 | 6.6 | 8.1 | 8.6 | 9.6 |
| 1800 | 6.1 | 3.8 | 4.4 | 4.7 | 6.0 | 7.6 | 8.0 | 9.0 |
| 1850 | 5.6 | 3.4 | 4.0 | 4.3 | 5.6 | 7.0 | 7.5 | 8.4 |
| 1900 | 4.9 | 2.9 | 3.4 | 3.7 | 4.8 | 6.2 | 6.6 | 7.5 |

\* thickness of an H$_2$O layer if it were spread evenly over the Martian surface in kilometers, referred to as a global equivalent layer (GEL)

**Table S7. Bulk H$_2$O storage capacity in the solid Martian mantle with the "LK19 (EH+L)" composition (m GEL*)**

| $T_p$ | best-fit | 1st | 5th | 10th | 50th percentile | 90th | 95th | 99th |
|:---:|:---:|:---:|:---:|:---:|:---:|:---:|:---:|:---:|
| (K) | (mean) | | | | (median) | | | |
| 1600 | 8.6 | 5.8 | 6.5 | 6.9 | 8.5 | 10.3 | 10.9 | 12.1 |
| 1900 | 5.0 | 2.8 | 3.3 | 3.7 | 4.9 | 6.4 | 6.9 | 7.8 |

* thickness of an H$_2$O layer if it were spread evenly over the Martian surface in meters, referred to as a global equivalent layer (GEL)

**Table S8. Bulk H$_2$O storage capacity in the solid Martian mantle with the "LK19 (EL+H)" composition (km GEL\*)**

| $T_p$ | best-fit | 1st | 5th | 10th | 50th percentile | 90th | 95th | 99th |
|---|---|---|---|---|---|---|---|---|
| (K) | (mean) | | | | (median) | | | |
| 1600 | 7.5 | 5.2 | 5.8 | 6.1 | 7.4 | 8.9 | 9.3 | 10.2 |
| 1900 | 4.2 | 2.4 | 2.9 | 3.2 | 4.2 | 5.4 | 5.8 | 6.5 |

\* thickness of an H$_2$O layer if it were spread evenly over the Martian surface in kilometers, referred to as a global equivalent layer (GEL)

**Table S9. Bulk H$_2$O storage capacity in the solid Martian mantle with the "YM20" composition (km GEL*)**

| $T_p$ | best-fit | 1st | 5th | 10th | 50th percentile | 90th | 95th | 99th |
|---|---|---|---|---|---|---|---|---|
| (K) | (mean) | | | | (median) | | | |
| 1600 | 7.7 | 5.4 | 6.0 | 6.4 | 7.7 | 9.2 | 9.6 | 10.5 |
| 1900 | 4.4 | 2.6 | 3.0 | 3.3 | 4.3 | 5.5 | 5.8 | 6.6 |

* thickness of an H$_2$O layer if it were spread evenly over the Martian surface in kilometers, referred to as a global equivalent layer (GEL)

**Table S10. Bulk $H_2O$ storage capacity in the solid Martian mantle with the "K22" composition (km GEL\*)**

| $T_p$ | best-fit | 1st | 5th | 10th | 50th percentile | 90th | 95th | 99th |
|---|---|---|---|---|---|---|---|---|
| (K) | (mean) | | | | (median) | | | |
| 1600 | 7.5 | 5.3 | 5.9 | 6.2 | 7.4 | 8.8 | 9.3 | 10.1 |
| 1900 | 4.0 | 2.3 | 2.7 | 3.0 | 3.9 | 5.0 | 5.3 | 6.0 |

\* thickness of an $H_2O$ layer if it were spread evenly over the Martian surface in kilometers, referred to as a global equivalent layer (GEL)

**Table S11. Additional literature data[1,2] on the water storage capacity of olivine, an update to the compilation of Dong et al. (2021)**

| method | calibration | ref. | exp. | $P$ (GPa) | $T$ (K) | initial bulk $H_2O$ (wt%) | phases | Fe (mol fraction) | $H_2O$ in NAM (ppm wt) |
|---|---|---|---|---|---|---|---|---|---|
| FTIR | P82+B03 | Z04 | Fa0 | 0.3 | 1273 | a few drops | n.a. but saturation verified | 0 | 5 |
| FTIR | P82+B03 | Z04 | Fa0 | 0.3 | 1323 | a few drops | n.a. but saturation verified | 0 | 6 |
| FTIR | P82+B03 | Z04 | Fa0 | 0.3 | 1373 | a few drops | n.a. but saturation verified | 0 | 7 |
| FTIR | P82+B03 | Z04 | Fa0 | 0.3 | 1423 | a few drops | n.a. but saturation verified | 0 | 8 |
| FTIR | P82+B03 | Z04 | Fa0 | 0.3 | 1473 | a few drops | n.a. but saturation verified | 0 | 9 |
| FTIR | P82+B03 | Z04 | Fa0 | 0.3 | 1523 | a few drops | n.a. but saturation verified | 0 | 10 |
| FTIR | P82+B03 | Z04 | Fa0 | 0.3 | 1573 | a few drops | n.a. but saturation verified | 0 | 12 |
| FTIR | P82+B03 | Z04 | Fa8.5 | 0.3 | 1273 | a few drops | n.a. but saturation verified | 0.085 | 14 |
| FTIR | P82+B03 | Z04 | Fa8.5 | 0.3 | 1323 | a few drops | n.a. but saturation verified | 0.085 | 16 |
| FTIR | P82+B03 | Z04 | Fa8.5 | 0.3 | 1373 | a few drops | n.a. but saturation verified | 0.085 | 20 |
| FTIR | P82+B03 | Z04 | Fa8.5 | 0.3 | 1423 | a few drops | n.a. but saturation verified | 0.085 | 24 |
| FTIR | P82+B03 | Z04 | Fa8.5 | 0.3 | 1473 | a few drops | n.a. but saturation verified | 0.085 | 28 |
| FTIR | P82+B03 | Z04 | Fa8.5 | 0.3 | 1523 | a few drops | n.a. but saturation verified | 0.085 | 29 |
| FTIR | P82+B03 | Z04 | Fa8.5 | 0.3 | 1573 | a few drops | n.a. but saturation verified | 0.085 | 35 |
| FTIR | P82+B03 | Z04 | Fa12 | 0.3 | 1273 | a few drops | n.a. but saturation verified | 0.12 | 20 |
| FTIR | P82+B03 | Z04 | Fa12 | 0.3 | 1323 | a few drops | n.a. but saturation verified | 0.12 | 28 |
| FTIR | P82+B03 | Z04 | Fa12 | 0.3 | 1373 | a few drops | n.a. but saturation verified | 0.12 | 29 |
| FTIR | P82+B03 | Z04 | Fa12 | 0.3 | 1423 | a few drops | n.a. but saturation verified | 0.12 | 30 |
| FTIR | P82+B03 | Z04 | Fa12 | 0.3 | 1473 | a few drops | n.a. but saturation verified | 0.12 | 37 |
| FTIR | P82+B03 | Z04 | Fa12 | 0.3 | 1523 | a few drops | n.a. but saturation verified | 0.12 | 43 |
| FTIR | P82+B03 | Z04 | Fa12 | 0.3 | 1573 | a few drops | n.a. but saturation verified | 0.149 | 46 |
| FTIR | P82+B03 | Z04 | Fa14.9 | 0.3 | 1273 | a few drops | n.a. but saturation verified | 0.149 | 24 |
| FTIR | P82+B03 | Z04 | Fa14.9 | 0.3 | 1323 | a few drops | n.a. but saturation verified | 0.149 | 31 |
| FTIR | P82+B03 | Z04 | Fa14.9 | 0.3 | 1373 | a few drops | n.a. but saturation verified | 0.149 | 32 |
| FTIR | P82+B03 | Z04 | Fa14.9 | 0.3 | 1423 | a few drops | n.a. but saturation verified | 0.149 | 42 |
| FTIR | P82+B03 | Z04 | Fa14.9 | 0.3 | 1473 | a few drops | n.a. but saturation verified | 0.149 | 43 |
| FTIR | P82+B03 | Z04 | Fa14.9 | 0.3 | 1523 | a few drops | n.a. but saturation verified | 0.149 | 48 |
| FTIR | P82+B03 | Z04 | Fa14.9 | 0.3 | 1573 | a few drops | n.a. but saturation verified | 0.149 | 56 |
| FTIR | P82+B03 | Z04 | Fa15.3 | 0.3 | 1273 | a few drops | n.a. but saturation verified | 0.153 | 24 |
| FTIR | P82+B03 | Z04 | Fa15.3 | 0.3 | 1323 | a few drops | n.a. but saturation verified | 0.153 | 34 |
| FTIR | P82+B03 | Z04 | Fa15.3 | 0.3 | 1373 | a few drops | n.a. but saturation verified | 0.153 | 34 |
| FTIR | P82+B03 | Z04 | Fa15.3 | 0.3 | 1423 | a few drops | n.a. but saturation verified | 0.153 | 41 |
| FTIR | P82+B03 | Z04 | Fa15.3 | 0.3 | 1473 | a few drops | n.a. but saturation verified | 0.153 | 46 |
| FTIR | P82+B03 | Z04 | Fa15.3 | 0.3 | 1523 | a few drops | n.a. but saturation verified | 0.153 | 48 |
| FTIR | P82+B03 | Z04 | Fa15.3 | 0.3 | 1573 | a few drops | n.a. but saturation verified | 0.153 | 54 |
| FTIR | P82+B03 | Z04 | Fa16.8 | 0.3 | 1273 | a few drops | n.a. but saturation verified | 0.168 | 27 |
| FTIR | P82+B03 | Z04 | Fa16.8 | 0.3 | 1323 | a few drops | n.a. but saturation verified | 0.168 | 36 |
| FTIR | P82+B03 | Z04 | Fa16.8 | 0.3 | 1373 | a few drops | n.a. but saturation verified | 0.168 | 40 |
| FTIR | P82+B03 | Z04 | Fa16.8 | 0.3 | 1423 | a few drops | n.a. but saturation verified | 0.168 | 50 |
| FTIR | P82+B03 | Z04 | Fa16.8 | 0.3 | 1473 | a few drops | n.a. but saturation verified | 0.168 | 55 |
| FTIR | P82+B03 | Z04 | Fa16.8 | 0.3 | 1523 | a few drops | n.a. but saturation verified | 0.168 | 58 |
| FTIR | P82+B03 | Z04 | Fa16.8 | 0.3 | 1573 | a few drops | n.a. but saturation verified | 0.168 | 64 |
| FTIR | P82+B03 | Z04 | Fa16.9 | 0.3 | 1273 | a few drops | n.a. but saturation verified | 0.169 | 26 |
| FTIR | P82+B03 | Z04 | Fa16.9 | 0.3 | 1323 | a few drops | n.a. but saturation verified | 0.169 | 36 |
| FTIR | P82+B03 | Z04 | Fa16.9 | 0.3 | 1423 | a few drops | n.a. but saturation verified | 0.169 | 48 |
| FTIR | P82+B03 | Z04 | Fa16.9 | 0.3 | 1473 | a few drops | n.a. but saturation verified | 0.169 | 57 |
| FTIR | P82+B03 | Z04 | Fa16.9 | 0.3 | 1523 | a few drops | n.a. but saturation verified | 0.169 | 59 |
| FTIR | P82+B03 | Z04 | Fa16.9 | 0.3 | 1573 | a few drops | n.a. but saturation verified | 0.169 | 65 |
| FTIR | B03 | G07 | ALB5 | 1.5 | 1568 | 7 | ol+en+sp+melt | 0 | 22 |
| FTIR | B03 | G07 | ALB6 | 1.5 | 1593 | 7 | fo+en+melt | 0 | 23 |
| FTIR | B03 | G07 | ALB8 | 2 | 1612 | 7 | fo+en+melt | 0 | 36 |
| FTIR | B03 | G07 | ALB10 | 1 | 1593 | 7 | fo+sp+melt | 0 | 37 |

1. <u>Data used and included in Table S11:</u>
   P82: Paterson, M. (1982). The determination of hydroxyl by infrared absorption in quartz, silicate glasses and similar materials. *Bulletin de Minéralogie*, **105**(1), 20–29; B03: Bell, D. R., Rossman, G. R., Maldener, J., Endisch, D., & Rauch, F. (2003). Hydroxide in olivine: A quantitative determination of the absolute amount and calibration of the IR spectrum.

**Table S12. Additional literature data[1,2] on the water storage capacity of wadsleyite, an update to the compilation of Dong et al. (2021)**

| method | calibration | ref. | exp. | $P$ (GPa) | $T$ (K) | initial bulk $H_2O$ (wt%) | phases | Fe (mol fraction) | $H_2O$ in NAM (wt%) |
|--------|------------|------|------|-----------|---------|---------------------------|--------|-------------------|---------------------|
| FTIR | B18 | F21 | S7114Ol | 17.5 | 2100 | 15 | wd+melt | 0.042 | 0.65 |
| FTIR | B18 | F21 | H4817Ol | 17.5 | 1700 | 15 | wd+en+melt | 0.057 | 1.35 |
| FTIR | B18 | F21 | H4898Ol | 17.5 | 1500 | 15 | wd+PhB+melt | 0.066 | 2.17 |
| FTIR | B18 | F21 | H4841Ol | 17.5 | 1900 | 15 | wd+melt | 0.051 | 1.02 |
| FTIR | B18 | F21 | H4821Ol | 17.5 | 1900 | 15 | wd+en+melt | 0.124 | 0.94 |
| FTIR | B18 | F21 | H4821Fo | 17.5 | 1900 | 15 | wd+en+melt | 0.000 | 0.73 |
| FTIR | B18 | F21 | H4898Fo | 17.5 | 1500 | 15 | wd+PhB+melt | 0.000 | 2.305 |
| FTIR | B18 | F21 | H4790Fo | 21 | 2000 | 5 | wd+en+melt | 0.000 | 0.58 |

1. Data used and included in Table S12:
   B18: Bolfan-Casanova, N., Schiavi, F., Novella, D., Bureau, H., Raepsaet, C., Khodja, H., & Demouchy, S. (2018). Examination of water quantification and incorporation in Transition Zone Minerals: wadsleyite, ringwoodite and phase D using ERDA (Elastic recoil detection Analysis). *Frontiers in Earth Science*, **6**, 75; F21: Fei, H., & Katsura, T. (2021). Water solubility in Fe-bearing wadsleyite at mantle transition zone temperatures. *Geophysical Research Letters*, 48, e2021GL092836.

2. Data used but **NOT** included in Table S12 (cf. Dong et al., 2021):
   D05: Demouchy, S., Deloule, E., Frost, D. J., & Keppler, H. (2005). Pressure and temperature dependence of water solubility in Fe-free wadsleyite. *American Mineralogist*, **90**, 1084–1091; I10: Inoue, T., Wada, T., Sasaki, R., & Yurimoto, H. (2010). Water partitioning in the Earth's mantle. *Physics of the Earth and Planetary Interiors*, **183**, 245–251; J05: Jacobsen, S. D., Demouchy, S., Frost, D. J., Ballaran, T. B., & Kung, J. (2005). A systematic study of OH in hydrous wadsleyite from polarized FTIR spectroscopy and single-crystal X-ray diffraction: Oxygen sites for hydrogen storage in Earth's interior. *American Mineralogist*, **90**, 61–70; K96: Kawamoto, T. (1996). Experimental constraints on differentiation and $H_2O$ abundance of calc-alkaline magmas. *Earth and Planetary Science Letters*, **144**, 577–589; L11: Litasov, K. D., Shatskiy, A., Ohtani, E., & Katsura, T. (2011). Systematic study of hydrogen incorporation into Fe-free wadsleyite. *Physics and Chemistry of Minerals*, **38**, 75–84; L03: Litasov, K., & Ohtani, E. (2003). Stability of various hydrous phases in CMAS pyrolite-$H_2O$ system up to 25 GPa. *Physics and Chemistry of Minerals*, **30**, 147–156; L08: Litasov, K., & Ohtani, E. (2008). Systematic study of hydrogen incorporation into Fe-bearing wadsleyite and water storage capacity of the transition zone. *AIP Conference Proceedings*, **987**, 113–118; S97: Smyth, J. R., & Kawamoto, T. (1997). Wadsleyite II: A new high pressure hydrous phase in the peridotite-$H_2O$ system. *Earth and Planetary Science Letters*, **146**,

**Table S13. Additional literature data[1,2] on the water storage capacity of ringwoodite, an update to the compilation of Dong et al. (2021)**

| method | calibration | ref. | exp. | $P$ (GPa) | $T$ (K) | initial bulk $H_2O$ (wt%) | phases | Fe (mol fraction) | $H_2O$ in NAM (wt%) |
|--------|-------------|------|------|-----------|---------|---------------------------|--------|-------------------|---------------------|
| FTIR | B18 | F20 | H4720 | 23 | 1600 | 15 | rw+aki+melt+st | 0.066 | 1.09 |
| FTIR | B18 | F20 | H4784L | 23 | 1800 | 5 | rw+melt | 0 | 1.12 |
| FTIR | B18 | F20 | H4800 | 23 | 1800 | 5 | rw+aki+melt | 0 | 1.71 |
| FTIR | B18 | F20 | H4784H | 23 | 1800 | 15 | rw+aki+melt | 0 | 1.74 |
| FTIR | B18 | F20 | H4723 | 23 | 1800 | 15 | rw+aki+melt | 0.0584 | 0.96 |
| FTIR | B18 | F20 | S6985 | 23 | 1800 | 5 | rw+aki+melt | 0.08 | 0.31 |
| FTIR | B18 | F20 | H4775 | 23 | 2000 | 5 | rw+melt | 0 | 0.49 |
| FTIR | B18 | F20 | H4793 | 23 | 2000 | 5 | rw+aki+melt | 0 | 0.89 |
| FTIR | B18 | F20 | H4805 | 23 | 2000 | 15 | rw+melt | 0 | 0.87 |
| FTIR | B18 | F20 | S7051 | 23 | 2000 | 22 | rw+aki?+melt | 0.0322 | 0.81 |
| FTIR | B18 | F20 | S7011 | 23 | 2000 | 15 | rw+aki+melt | 0.0501 | 0.84 |
| FTIR | B18 | F20 | H4711 | 23 | 2000 | 5 | rw+aki+melt | 0.0663 | 0.28 |
| FTIR | B18 | F20 | H4698 | 23 | 2000 | 15 | rw+aki+melt | 0.0682 | 0.87 |
| FTIR | B18 | F20 | H4754 | 23 | 2000 | 1.5 | rw+aki+melt | 0.0845 | 0.18 |
| FTIR | B18 | F20 | H4762 | 23 | 2000 | 15 | rw+melt+st | 0.1109 | 0.84 |

1. Data used and included in Table S13:
   B18: Bolfan-Casanova, N., Schiavi, F., Novella, D., Bureau, H., Raepsaet, C., Khodja, H., & Demouchy, S. (2018). Examination of water quantification and incorporation in Transition Zone Minerals: wadsleyite, ringwoodite and phase D using ERDA (Elastic recoil detection Analysis). *Frontiers in Earth Science*, **6**, 75; F20: Fei, H., & Katsura, T. (2020). Pressure dependence of proton incorporation and water solubility in olivine. *Journal of Geophysical Research: Solid Earth*, **125**(2), e2019JB018813.

2. Data used but **NOT** included in Table S13 (cf. Dong et al., 2021):
   I10: Inoue, T., Wada, T., Sasaki, R., & Yurimoto, H. (2010). Water partitioning in the Earth's mantle. *Physics of the Earth and Planetary Interiors*, **183**, 245–251; O00: Ohtani, E., Mizobata, H., & Yurimoto, H. (2000). Stability of dense hydrous magnesium silicate phases in the systems $Mg_2SiO_4$-$H_2O$ and $MgSiO_3$-$H_2O$ at pressures up to 27 GPa. *Physics of the Earth and Planetary Interiors*, **27**, 533–544; S03: Smyth, J. R. et al. (2003). Structural systematics of hydrous ringwoodite and water in Earth's interior. *American Mineralogist*, **88**, 1402–1407.